\documentclass{aa}  

\usepackage{natbib}
\bibpunct{(}{)}{;}{a}{}{,} 

\usepackage{pdflscape}

\usepackage{color}
\usepackage{times}
\usepackage{soul}
\usepackage{amsmath,amssymb,amsfonts,txfonts}
\usepackage[colorinlistoftodos]{todonotes}
\usepackage{multirow}
\usepackage{nicefrac}
\usepackage{notes2bib}
\usepackage{tabularx}

\usepackage{hyperref}  
\hypersetup{colorlinks=true,linkcolor=[rgb]{1.,0.2,0.2},citecolor=[rgb]{0.1,0.4,1.},filecolor=[rgb]{0.7,0.2,0.2},urlcolor=[rgb]{0.7,0.2,0.2}}

\usepackage{graphicx}

\newcommand {\sn} {$\left<S/N\right>\,$}

\newcommand {\vel}{$\,{\rm km}\,{\rm s}^{-1}$}
\newcommand {\ma} {{MACS\,1206}}
\newcommand {\mb} {{MACS\,0416}}
\newcommand {\ab} {{AS1063}}
\newcommand {\bo} {{B18}}
\newcommand {\cama} {{C17b}}
\newcommand {\camb} {{C17a}}
\newcommand {\caab} {{C16}}

\definecolor{darkspringgreen}{rgb}{0.09, 0.45, 0.27}

\begin{document} 

\title{Enhanced cluster lensing models with measured galaxy kinematics}

\authorrunning{Bergamini et al}

\author{P.~Bergamini  \inst{\ref{inafpd},\ref{inafna}\thanks{E-mail: \href{mailto:pietro.bergamini@phd.unipd.it}{pietro.bergamini@phd.unipd.it}}}
 \and
P.~Rosati  \inst{\ref{unife},\,\ref{inafbo}} \and
A.~Mercurio \inst{\ref{inafna}}       \and
C.~Grillo  \inst{\ref{unimilano},\,\ref{dark}}  \and
G.~B.~Caminha       \inst{\ref{Kapteyn}}  \and
M.~Meneghetti   \inst{\ref{inafbo},\,\ref{infnbologna}} \and
A.~Agnello       \inst{\ref{dark}}  \and
A.~Biviano       \inst{\ref{inaftrieste}}  \and
F.~Calura   \inst{\ref{inafbo}} \and
C.~Giocoli   \inst{\ref{unife}, \ref{inafbo}, \ref{infnbologna}} \and
M.~Lombardi  \inst{\ref{unimilano}}  \and
G.~Rodighiero  \inst{\ref{inafpd}}\and
E.~Vanzella       \inst{\ref{inafbo}}  \
 }
   
\institute{
Dipartimento di Fisica e Astronomia "G. Galilei", Universit\`a di Padova, Vicolo dell’Osservatorio 3, I-35122, Italy\label{inafpd}
\and Dipartimento di Fisica e Scienze della Terra, Universit\`a degli Studi di Ferrara, Via Saragat 1, I-44122 Ferrara, Italy\label{unife}\and
INAF - Osservatorio Astronomico di Capodimonte, Via Moiariello 16, I-80131 Napoli, Italy\label{inafna} \and
Dipartimento di Fisica, Universit\`a  degli Studi di Milano, via Celoria 16, I-20133 Milano, Italy\label{unimilano} \and
DARK, Niels-Bohr Institute, Lyngbyvej 2, 2100 Copenhagen\label{dark}\and
INAF - OAS, Osservatorio di Astrofisica e Scienza dello Spazio di Bologna, via Gobetti 93/3, I-40129 Bologna, Italy\label{inafbo}\and
INFN, Sezione di Bologna, viale Berti Pichat 6/2, 40127 Bologna, Italy\label{infnbologna} \and 
INAF - Osservatorio Astronomico di Trieste, via G. B. Tiepolo 11, I-34131, Trieste, Italy\label{inaftrieste} \and 
Kapteyn Astronomical Institute, University of Groningen, Postbus 800, 9700 AV Groningen, The Netherlands \label{Kapteyn} 
             }

\date{Received May 28, 2019; accepted XXX}

\abstract
  {
We present an improved determination of the total mass distribution of three CLASH/Hubble Frontier Fields massive clusters, MACS~J1206.2$-$0847 ($z=0.44$), MACS~J0416.1$-$2403 ($z=0.40$), Abell~S1063 ($z=0.35$). We specifically reconstruct the sub-halo mass component with robust stellar kinematics information of cluster galaxies, in combination with precise strong lensing models based on large samples of spectroscopically identified multiple images. We use VLT/MUSE integral-field spectroscopy in the cluster cores to measure the stellar velocity dispersion, $\sigma$, of 40-60 member galaxies per cluster, covering 4-5 magnitudes to $m_{F160W}\simeq 21.5$. We verify the robustness and quantify the accuracy of the velocity dispersion measurements with extensive spectral simulations. With these data, we determine the normalization and slope of the galaxy $L\mbox{-}\sigma$ Faber-Jackson relation in each cluster and use these parameters as a prior for the scaling relations of the sub-halo population in the mass distribution modeling. When compared to our previous lens models, the inclusion of member galaxies' kinematics provides a similar precision in reproducing the positions of the multiple images. However, the inherent degeneracy between the central effective velocity dispersion, $\sigma_0$, and truncation radius, $r_{cut}$, of sub-halos is strongly reduced, thus significantly alleviating possible systematics in the measurements of sub-halo masses. The three independent determinations of the $\sigma_0\mbox{-}r_{cut}$ scaling relation in each cluster are found to be fully consistent, enabling a statistical determination of sub-halo sizes as a function of $\sigma_0$, or halo masses. Finally, we derive the galaxy central velocity dispersion functions of the three clusters projected within 16\% of their virial radius, finding that they are well in agreement with each other. We argue that such a methodology, when applied to high-quality kinematics and strong lensing data, allows the sub-halo mass functions to be determined and compared with those obtained from cosmological simulations.  

}


   \maketitle


\section{Introduction}
Strong gravitational lensing is a powerful technique to probe the total projected mass distribution of the inner regions of galaxy clusters and, therefore, the dark matter (DM) distribution once the baryonic mass components are independently mapped. By characterizing the substructure of cluster cores on different scales and the mass density profile of the innermost regions, one can test the $\Lambda$CDM structure formation scenario and indirectly constrain DM physical properties by comparing reconstructed mass maps with the latest cosmological simulations (e.g., \citealt{Diemand_2011}).  
In recent years, dedicated  Hubble Space Telescope (HST) observational campaigns, such as the Cluster Lensing And Supernova survey with Hubble (CLASH, \citealt{Postman_2012_clash}), the Hubble Frontier Field campaign (HFF, \citealt{Lotz_HST}), and the Reionization Lensing Cluster Survey (RELICS, \citealt{Coe_2019}) have provided multi-band data with unprecedented quality on a sizable set of massive galaxy clusters. In parallel, extensive spectroscopic campaigns with highly-multiplexing multi-slit instruments (e.g., CLASH-VLT, Rosati et al. in prep., \citealt{Balestra_MACS0416}), and especially with the MUSE (Multi Unit Spectroscopic Explorer) integral field spectrograph on the VLT (Very Large Telescope) \citep{Bacon_MUSE}, have secured redshifts for hundreds of multiple lensed images in cluster cores, as well identified large samples of cluster galaxies.  The combination of these new imaging and spectroscopic data sets have enabled the development of new high-precision parametric strong lensing models based on 50-100  bona-fide multiple images per cluster and highly complete samples of cluster members (e.g., \citealt{Richard_2014}, \citealt{Grillo_2015}, \citealt{Jauzac_2015}, \citealt{Limousin_2016}, \citealt{Kawamata_2016}, \citealt{Lagattuta_2017}, \citealt{Caminha_rxc2248}, \citealt{Caminha_macs0416}, \citealt{Caminha_macs1206}, \citealt{Bonamigo_2018}, \citealt{Caminha_2019}). These new lens studies typically reach a root mean square difference between the observed and predicted positions of multiple images on the lens plane of $\Delta_{rms}\approx 0.5\arcsec$. The latter is commonly used as a simple figure of merit for model accuracy. Nonetheless, the relatively large number of parameters used to describe the cluster mass distribution still suffer from strong internal degeneracies. The generally complex cluster mass distribution in parametric lens models is generally separated into a cluster-scale diffuse component, made of one or more large halos, and a clumpy distribution traced by cluster galaxies \citep{Natarajan_1997}. The latter describes the sub-halo population of DM halos \citep{delucia04,giocoli10a} and hence the inner substructure of the cluster mass distribution \citep{springel01}. In the effort to limit the number of free parameters, this sub-halo component is generally modeled adopting two scaling relations, which link the internal velocity dispersion and size of each halo with the luminosity of member galaxies. A fixed mass-to-light scaling is thereby assumed for all sub-halos. Since strong lensing models constrain the total projected mass within each family of multiple image positions, a certain amount of degeneracy is always present between the velocity dispersion, the size (or equivalently the profile), and the shape of each halo component of the mass distribution. Given the cross-talks among different halo components in strong lensing models, such a degeneracy can lead to systematics in the reconstruction of the substructure at different scales, for example with a transfer of mass between the clumpy and diffuse mass components.  As we demonstrate in this work, such a degeneracy can be broken and significantly reduced, on the scale of the sub-halos, by using an independent measurement of the internal stellar velocity dispersion of cluster members.

A combination of lensing and kinematic measurements has long been exploited in the study of the mass density profile of field early-type galaxies. This combination has proved to be particularly effective since the two diagnostics complement each other (e.g., \citealt{Treu_2004}; \citealt{Czoske_2008}; \citealt{Barnabe_2009}), breaking the mass-anisotropy and mass-sheet (e.g., \citealt{Falco_1985}; \citealt{Grav_lens_2006}) degeneracies of the dynamical and lensing analyses, respectively. Joint strong lensing and stellar-dynamical studies have been used to determine the average logarithmic density slope of the total mass inside the Einstein radius (\citealt{Treu_2004}; \citealt{Koopmans_2009}) and to decompose the total mass distribution into luminous and dark components of the lens galaxies belonging to the Lenses Structure and Dynamics (LSD) and Sloan Lens ACS (SLACS) surveys. These two surveys have measured a remarkably homogeneous total (luminous and dark) mass density profile that is consistent with an isothermal one (i.e., $\rho \propto 1/r^{2}$) out to a few hundreds of kiloparsecs (see \citealt{Gavazzi_2007}; \citealt{Bolton_2008}). A one-component isothermal model is fully characterized by the value of an effective velocity dispersion, approximated within $\lesssim 3\%$ by the value of the galaxy central stellar velocity dispersion (i.e., the velocity dispersion of the stars projected within a disk of radius $R_{e}/8$; see, e.g., \citealt{Treu_2006}; \citealt{Bolton_2008}). This result is theoretically supported by the Jeans equation for realistic stellar density distributions (e.g., \citealt{Jaffe_1983}; \citealt{Hernquist_1990}) embedded in a globally isothermal distribution (see \citealt{Kochanek_1993K}), as observed in samples of nearby and luminous early-type galaxies (e.g., \citealt{Kochanek_1994}; \citealt{Grillo_2008}).

A combination of the spatially resolved kinematics of the brightest cluster galaxy (BCG) and strong lensing modeling in the inner cores of massive clusters has also been used to constrain the central slope of their mass density profiles using simplified dynamical models (\citealt{Sand_2004}, \citealt{Newman_2013}). 
A first attempt to include cluster galaxy kinematics in cluster strong lensing models was made by \cite{Verdugo_2007}. A more extensive study was carried out by \cite{Monna_2015} in Abell 383 at $z=0.187$ with early CLASH data. They integrated the velocity dispersion measurements of 21 member galaxies, obtained with the Hectospec fiber spectrograph at the MMT, into a lens model based on the identification of  nine multiply imaged systems (six of which spectroscopically confirmed). Thus, they show how typical degeneracies among lens model parameters describing the galaxy mass-to-light scaling relation can be significantly reduced. In particular, a meaningful scaling relation between galaxy halo truncation radii and velocity dispersions can be derived (see also \citealt{Monna_2017b}). Additional constraints on halo sizes can be obtained by modeling the surface brightness distribution of strong lensing features around individual member galaxies (see also \citealt{Suyu_2010}, \citealt{Eichner_2013} who did not use galaxy kinematics).

In our work, we extend the methodology introduced by \cite{Monna_2015}, by using a much improved HST and spectroscopic data set on three CLASH/HFF clusters, specifically an extended set of velocity dispersion measurements obtained with the MUSE spectrograph at the VLT. A robust characterization of cluster sub-halo populations, particularly the distribution of their sizes and masses, as well as their abundance, offer a critical test of the predictions of cosmological simulations (e.g., \citealt{Limousin_2009}, \citealt{Grillo_2015}, \citealt{Munari_2016}, \citealt{Natarajan_2009}, \citeyear{Natarajan_2017}). Such a comparison can shed light on baryonic processes shaping cluster substructure, dynamical processes leading to halo stripping in different environments, and indirectly on the nature of dark matter \citep{despali17b,chua17,nipoti18,niemiec19}.

This work is organized as follows. In Sec.\,\ref{sec.:Photometric and spectroscopic data}, we describe our imaging and spectroscopic data sets. In Sec.\, \ref{sec.:Cluster members: spectral extraction and internal kinematics}, we detail how internal velocity dispersions of member galaxies are measured, including spectral simulations to assess their robustness. Strong lensing models for the three clusters under study are described in Sec.\,\ref{sec.:lensing_models}, while  the specific methodology to incorporate galaxy kinematics information into our lens models is discussed in Sec.\,\ref{sec.:Lensing_kinematics}. Results are discussed in Sec.\,\ref{Results} where we also present the velocity dispersion functions for the three clusters. In Sec.\,\ref{Conclusions}, we summarize the main conclusions of our study.\\
Throughout this article, we adopt a flat $\mathrm{\Lambda CDM}$ cosmology with $\mathrm{\Omega_m = 0.3}$ and $\mathrm{H_0 = 70\,km\,s^{-1}Mpc^{-1}}$. With these parameters, 1\arcsec\ corresponds to a physical scale of 5.68, 5.34 and 4.92 kpc at $z=0.439,0.396,0.348$, respectively the redshift of the three clusters of this study: \ma, \mb\ and \ab\ (see below).
All magnitudes refer to the AB system.

\section{Photometric and spectroscopic data}
\label{sec.:Photometric and spectroscopic data}
This section summarizes the photometric and spectroscopic data sets for the three clusters used in this work, namely \object{MACS~J1206.2$-$0847}, \object{MACS~J0416.1$-$2403}, and \object{Abell S1063} (a.k.a. RXJ~2248.7$-$4431), hereafter \ma, \mb\ and \ab, at redshifts 0.439, 0.396 and 0.348, respectively (see Table\,\ref{table:cluster_sample}).
These clusters were observed with HST in 16 broad band filters, from UV to near-IR, as part of the CLASH program. HST imaging of \mb\ and \ab\ was significantly augmented with the HFF program by adding deep exposures in seven filters (F435, F606W, F814W, F105W, F125W, F140W, F160W). The three clusters were also part of an extensive spectroscopic campaign with the CLASH-VLT Large program (P.I. P. Rosati), using the VIMOS high-multiplexing spectrograph, which provided over 4000 redshifts in each of the three clusters, over an area of $\sim\! 25\times 25\, \rm{arcmin}^2$. These data sets yielded approximately 600 spectroscopic members for \ma\ (\citealt{Biviano_2013_macs1206}, \citealt{Girardi_2015}), 900 members for \mb\ \citep{Balestra_MACS0416} and over 1200 members for \ab\ (Mercurio et al. 2019, in prep.). Spectroscopic information in the cores of the three clusters has been significantly enhanced with the MUSE integral field spectrograph at the VLT, which is at the basis of the kinematic measurements presented in this work. These data, which are described in more detail below, have enabled new high-precision strong lensing models based on large samples of multiply lensed sources (see \citealt{Caminha_macs0416}, \citealt{Caminha_macs1206}, \citealt{Caminha_rxc2248} and Sec.\,\ref{sec.:lensing_models}). MUSE has a field of view of 1 arcmin$^2$, a spatial sampling of 0.2\arcsec, a spectral resolution of $\sim\! 2.4$\,\AA\ over the spectral range $4750\,\mbox{-}\,9350\,$\AA, with a spectral sampling of 1.25\,\AA/pix. \smallskip

\noindent {\bf \ma}:\, MUSE data were obtained between 2015 and 2016\footnote{ID 095.A-0181(A) and 097.A-0269(A) (P.I. J. Richard)}, the redshift measurements of the member galaxies and a large sample of multiple images were presented in \cite{Caminha_macs1206} (hereafter \cama). Three MUSE pointings cover a total area of $2.63 \, {\rm arcmin}^2$, mapping the SE--NW elongation of the cluster. The exposure time is 8.5 hours in the central $\sim 0.5 \,{\rm arcmin}^2$ and 4 hours in the remaining area. \smallskip

\noindent {\bf \mb}:\, MUSE archival observations used in this work were presented in \cite{Caminha_macs0416} (hereafter \camb), along with the redshift catalog, and consist of two pointings\footnote{ID 094.A-0115B (PI J. Richard) and 094.A0525(A) (PI F.E. Bauer)}, one of 2 hours in the NE region, with a seeing of $0.5\arcsec$, and a SW pointing of 11 hours, with a seeing of $1\arcsec$. \smallskip

\noindent {\bf \ab}:\,  MUSE data consist of two pointings, which were presented in \cite{Karman_2015} and \cite{Karman_2017}. The SW pointing\footnote{ID 60.A-9345 (P.I.: K.~Caputi\, \&\, C.~Grillo)} has an exposure of 3.1 hours and seeing $\sim 1.1\arcsec$, the NE pointing\footnote{ID 095.A-0653 (P.I. K.~Caputi)} has an exposure of 4.8 hours and seeing of $0.9\arcsec$.

\section{Cluster members: spectral extraction and internal kinematics}
\label{sec.:Cluster members: spectral extraction and internal kinematics}
In this section, we describe the methodology adopted to extract the spectra of the cluster members from the MUSE data-cubes and to measure their internal stellar velocity dispersions.
Catalogs of cluster members for \ma, \mb, \ab\ were presented in \cama, \camb\  and \cite{Caminha_rxc2248} (hereafter \caab), respectively, with the main objective of identifying the sub-halos to be included in the lens models (see below). Cluster members were defined as galaxies in the redshift intervals 0.425\,\mbox{-}\,0.453, 0.382\,\mbox{-}\,0.410, and 0.335\,\mbox{-}\,0.362 for the three clusters respectively, lying  within a rest-frame velocity of approximately $\pm 3000$\,\vel around the median redshift of the cluster.  With 114\,\mbox{-}\,145 spectroscopic members per cluster in the HST FoV, the extensive CLASH multi-band information was used to obtain highly ($\sim\! 95\%$) complete and pure samples of photometric members down to $m_{\rm F160W}=24$, following the method described in \cite{Grillo_2015}.

The spectra of cluster members are extracted from the MUSE data-cubes within apertures of $R_{ap}=0.8\arcsec$ radius. The latter is found to be a good compromise between the signal-to-noise of the spectra and the contamination from other sources in the field (either interlopers along the line-of-sight or close members). All the extractions are visually inspected to assess possible contamination from nearby sources and the apertures are reduced in specific cases down to 0.6\arcsec. The spectra in which the contamination from bright nearby members is too strong are discarded as the velocity dispersion of the fainter galaxy is likely biased. 

\begin{figure}[h!]
	\centering
	\includegraphics[width=9cm]{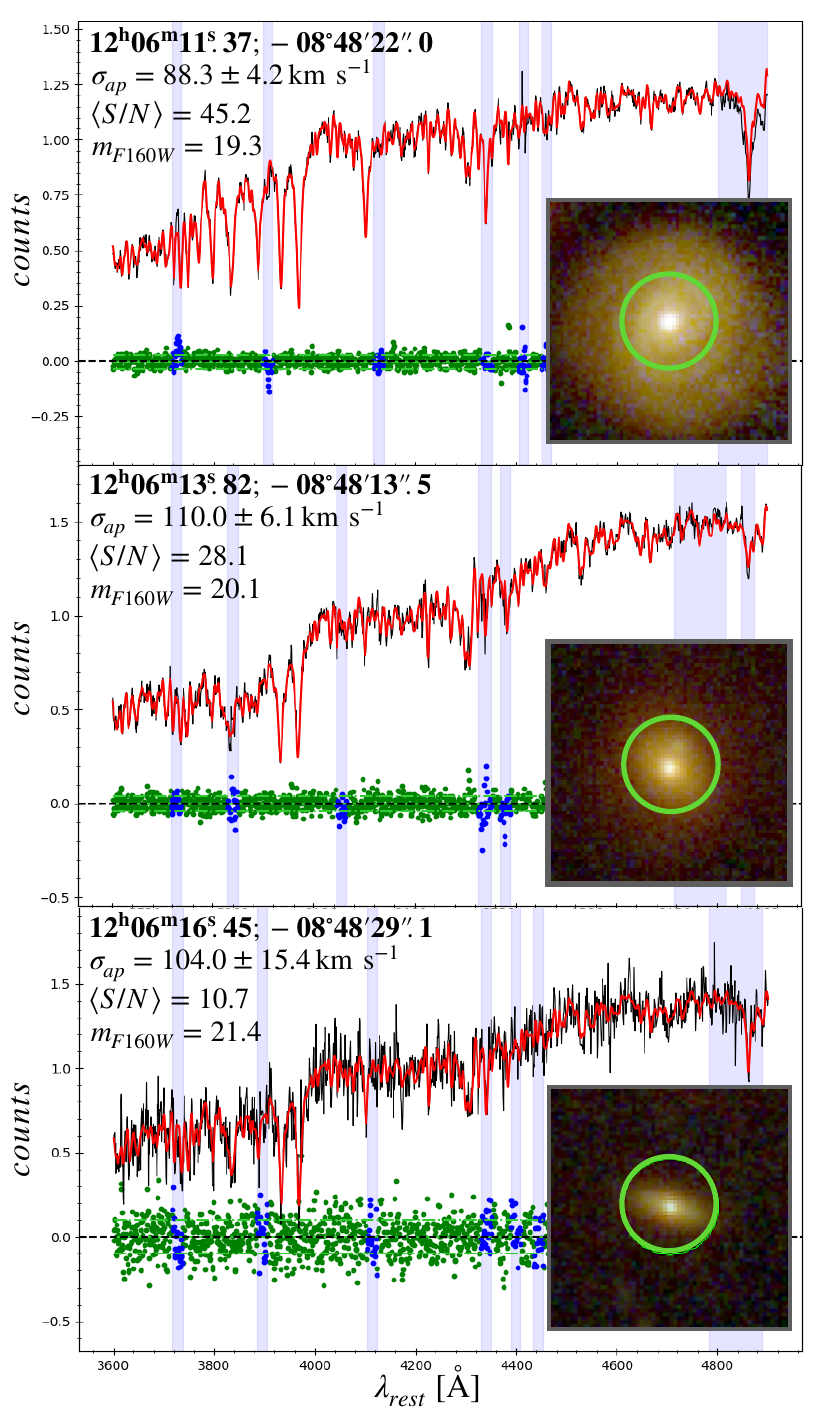}
	\caption{Results from line-of-sight velocity dispersion fitting of spectra of three cluster members in \ma\ as obtained with pPXF. 
	Galaxy spectra are shown in black; red curves are the pPXF best-fit models, while the green points correspond to the data$-$model residuals. The blue shaded regions along the wavelength axis were excluded in the fitting procedure due to the presence of noisy sky subtractions around emission lines in the spectra. Corresponding residuals in these regions are marked in blue.
	The first from the top is a high signal-to-noise E+A galaxy. The second is a passive galaxy spectrum with a $\left< S/N \right>=28.1$, corresponding approximately to the mean S/N of our galaxy sample. The bottom spectrum has a $\left< S/N \right>=10.7$, close to our lower limit for reliable velocity dispersion measurements.  Coordinates, measured velocity dispersion ($\sigma_{ap}$), mean signal-to-noise and F160W magnitudes are indicated in each panel. Cutouts are HST RGB images, $4$\arcsec\ across, showing in green the apertures of $0.8$\arcsec\ radius used for the spectral extraction.}
	\label{fig:pPXF_spectra}
\end{figure}

\begin{table*}[h!]   
	\tiny
	\def\arraystretch{1.6}
	\centering    
	\begin{tabular}{|c|c|c|c|c|c|c|}
	   \hline
	   \bf{Cluster} & $\bf{z}$ & $\bf{N_{m}^{meas}\ (N_{m}^{tot})}$ & $\bf{N_{im}\ (N_{fam})}$ &  $\bf{M_{200c}[10^{15} M_{\odot}]}$ & $\bf{R_{200c}[Mpc]}$ & $\bf{N_{m}(<0.16\,R_{200c})}$ \cr
	   \hline
	   \hline
	   \textbf{MACS J1206.2$\mathbf{-}$0847} & $0.439$ & $58\ (258)$  & $82\ (27)$ & $(1.59\pm 0.36)$ & $(2.06\pm 0.16)$  & $179$ \cr
	   \hline
	   \textbf{MACS J0416.1$\mathbf{-}$0403} & $0.396$ & $49\ (193)$  & $102\ (37)$ & $(1.04\pm 0.22)$ & $(1.82\pm 0.13)$ & $124$ \cr
	   \hline
	   \textbf{Abell S1063} & $0.348$ & $37\ (222)$  & $55\ (20)$ & $(2.03\pm 0.67)$ & $(2.32\pm 0.26)$ & $199$ \cr
	   \hline
	\end{tabular}\\[1ex]
    \caption{Most relevant parameters of three clusters of this work. Redshift ($z$), number of cluster members with measured velocity dispersion ($\mathrm{N_{m}^{meas}}$), total number of cluster members within the HST field with $m_{F160W}<24$ included in the lens models ($\mathrm{N_{m}^{tot}}$), number of spectroscopically confirmed multiple images ($\mathrm{N_{im}}$), number of image families ($\mathrm{N_{fam}}$), $\mathrm{M_{200c}}$ and $\mathrm{R_{200c}}$ values from \cite{Umetsu_2014}, and number of cluster members within a radius of $0.16\,\mathrm{R_{200c}}$ 
    (last column).
    }   
    \label{table:cluster_sample} 
\end{table*}

We measure the stellar line-of-sight velocity distribution (hereafter LOSVD) of cluster members using the public software pPXF (penalized pixel-fitting, \citealt{Cappellari_2004}), with the latest improvements included in the 02/2018 python version \citep{Cappellari_2017}. With pPXF, we determine the best-fit LOSVD parameters (first and higher moments) by performing a cross-correlation of the observed spectrum with a set of spectral templates.
The best-fit is obtained by minimizing a $\chi^2$ between the template and the observed spectrum, including an additional penalty function whose weight is regulated by a parameter $\lambda$. This bias parameter is used to suppress higher velocity moments (h3, h4), when they become unreliable in spectra with low velocity dispersion and low S/N (see \citealt{Cappellari_2004}). For each spectrum, we measure a mean signal-to-noise, \sn\!, over the selected wavelength range. We then adopt a relation between $\lambda$ and \sn\!, as suggested in \cite{Cappellari_2010}, which we tested with extensive spectral simulations described in Appendix~\ref{sec.:appendix_ppxf}.

To perform the pPXF spectral fits, we use a subset of 105 stellar templates of different spectral types, drawn from the National Optical Astronomy Observatory library \citep{stellar_templates}. To match the typical underlying stellar populations of early-type galaxies in the cluster cores, most of the templates are of G, K, M spectral classes. In addition, we include 10 A-stars to extend spectral fits to a non-negligible fraction of E+A galaxies in our cluster sample and a few O and B stars. The stellar templates cover a wavelength range from 3465\,\AA\ to 9469\,\AA, with a sampling of 0.4\,\AA/pix and have an intrinsic resolution of 1.35\,\AA\ full width at half-maximum.

In Appendix\,\ref{sec.:appendix_ppxf}, we describe in detail a set of simulations which reproduce our galaxy spectra with varying input LOSVD parameters, redshift  and \sn\!, to quantify the accuracy and precision of pPXF in measuring the velocity dispersions $\sigma$ of member galaxies, thus optimizing  pPXF input parameters. Simulated spectra are constructed from model spectra spanning the spectral type diversity of our cluster galaxy populations and by adding noise drawn from the variance map of the reduced MUSE data-cubes.

With these simulations, we check the reliability of the statistical error provided by pPXF on our data, as well as the presence of systematic errors as a function of \sn and input velocity dispersion ($\sigma_{\rm in}$). The latter can become important especially when measuring velocity dispersions of low mass galaxies, as we approach the MUSE instrument resolution. We find that statistical errors are generally underestimated by $\sim\! 20$\% for \sn$>$15, up to $\sim\! 25$\% for \sn$\sim\! 10$, whereas a positive bias of a few \vel\ becomes evident for $\sigma\lesssim 100$\,\vel\ at high \sn\!. In addition, we find that measurements become increasingly uncertain at \sn$<10$. We therefore include in our galaxy kinematic sample only galaxies with \sn$>10$ and $\sigma >80$\,\vel. In all cases, we use empirical formulas (see Appendix\,\ref{sec.:appendix_ppxf}) to correct the measured velocity dispersions and their uncertainties in different \sn and $\sigma$ regimes. \\
Simulations were also used to choose the optimal wavelength range for pPXF fits, for a given galaxy redshift, by excluding regions of low S/N due to the MUSE sensitivity curve, particularly on the red side of the spectrum strongly affected by sky lines residuals. The resulting selected rest-frame wavelength ranges are $3600\,\mbox{-}\,4900$\,\AA, $3600\,\mbox{-}\,5200$\,\AA\ and $3600\,\mbox{-}\,5300$\,\AA\ for \ma, \mb\ and \ab, respectively. The LOSVD input velocity value (zero moment) is taken from our redshift catalogs. The measured parameters are the velocity shift (V, typically within 50\,\vel), the velocity dispersion $\sigma$, and higher moments (h3, h4), and their one-standard deviation errors. 
All velocity dispersions measured with pPXF are labeled in the following with the subscript ``$ap$'' to emphasize that these are line-of-sight quantities within an aperture, as opposed to $\sigma_0,\,{\rm and}\,\sigma_{LT}$, which refer to parameters inferred with the lens models (see next session). 
Fig.\,\ref{fig:pPXF_spectra} shows examples of pPXF spectral fitting for three cluster galaxies in \ma, one in the high \sn regime, one for the median \sn of our sample ($\sim 30$), and one at the limiting \sn $\sim 10$.

In Table\,\ref{table:cluster_sample}, we quote the number of uncontaminated spectra extracted in each cluster ($\mathrm{N_m^{meas}}$), for which we can reliably measure velocity dispersions, together with the total number of spectro-photometric members and relevant cluster parameters. In Fig.\,\ref{fig:green_plots}, the data points correspond to measured velocity dispersions of member galaxies as a function of their F160W magnitude, defining the Faber-Jackson relation in the three clusters.

\section{Strong lensing models}
\label{sec.:lensing_models}
Accurate strong lensing models were developed for \ma, \mb\ and \ab\ in \cama, \camb\ and \caab, respectively. These models were further refined in \cite{Bonamigo_2018} (\bo\ hereafter), who included the mass distribution of the hot gas component in each cluster, as derived from the Chandra X-ray data, dominating the smooth baryonic cluster component. As customary in cluster strong lensing modeling, none of these models included any kinematic information on cluster galaxies. 
We describe here our methodology which combines the \bo\ lens models with internal stellar kinematics derived from a large number of velocity dispersions measured with MUSE. For each cluster, we use the same catalogs of multiple images and cluster galaxies as in \bo\ and the Caminha et al. models, and likewise we employ the public software {\texttt{LensTool}} (\citealt{Kneib_lenstool}, \citealt{Jullo_lenstool}, \citealt{Jullo_Kneib_lenstool}). \\
A parametric lens model for the total mass distribution of each cluster is optimized searching for the set of parameters $\pmb{\xi}$, which minimize the $\chi^2$ defined on the lens plane as \citep{Jullo_lenstool}:

\begin{equation}
    \label{eq.: chisq_lenstool}
    \chi^2(\pmb{\xi}) := \sum_{j=1}^{N_{fam}} \sum_{i=1}^{N_{im}^j} \left(\frac{\left\| \mathbf{x}_{i,j}^{obs} - \mathbf{x}_{i,j}^{pred} (\pmb{\xi}) \right\|}{\Delta x_{i,j}}\right)^2,
\end{equation}

\begin{figure}[h!]
	\centering
	\includegraphics[width=9.0cm]{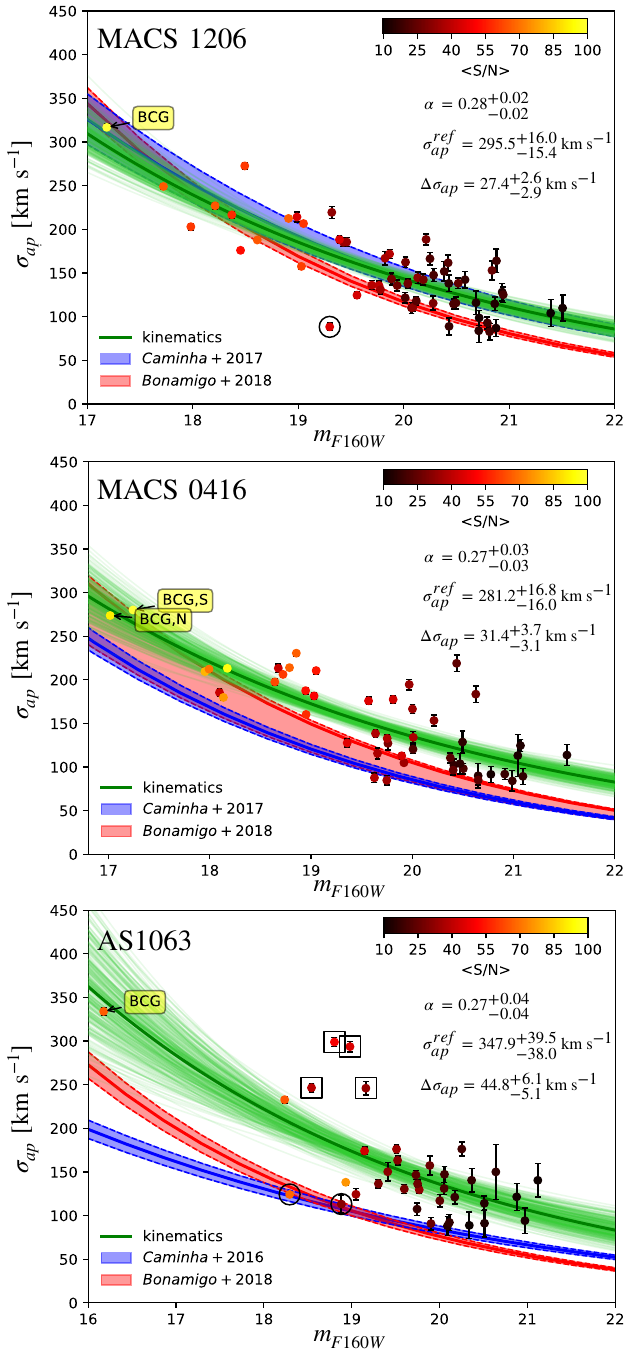}
	\caption{Data points are measured stellar velocity dispersions of cluster members, color-coded according to their spectral \sn. Green lines are 300 scaling relations randomly drawn from the posterior distributions of the $\sigma_{ap}\mbox{-}mag$ scaling relation parameters, $\alpha$, $\sigma_{ap}^{ref}$, $\Delta\sigma_{ap}$, obtained from fitting the data points (see Fig.\,\ref{fig:emcee_macs1206}). Optimized parameters are also quoted. Red and blue areas are obtained by projecting the 3D $\sigma\mbox{-}L$ scaling relations from our previous lens models with no kinematic prior (see text). Some velocity dispersion measurements, which deviate significantly from the scaling relations, are marked with squares and circles and discussed in the text.}
	\label{fig:green_plots}
\end{figure}

\noindent where $N_{im}^j$ identifies the number of multiple images associated to the same source $j$ (usually called a family),  $N_{fam}$ is the number of families, $ \mathbf{x}^{obs} $ are the observed positions of the multiple images on the lens plane, $\mathbf{x}^{pred}$ are their predicted positions, given the set of model parameters $\pmb{\xi}$, and $\Delta x_{i,j}$ represent the uncertainties on the observed positions. Following \bo, we perform a first optimization of our lensing models assuming a positional error of 0.5\arcsec\ for images identified in HST and 1\arcsec\ for those only found in the MUSE data. We multiply these errors by a constant factor ensuring that the best-fit $\chi^2$ is close to the number of degree of freedom of the models. These updated errors are then used to sample the posterior distributions of the free parameters.\\
Following the \bo\ modeling, the total mass distribution (or equivalently the gravitational potential $\phi$) of each cluster is described as the sum of three contributions: 1) an elliptical large-scale smooth halo, which is further decomposed in a DM component and a smooth gas mass component, both described as dual pseudo-isothermal elliptical density (dPIE) profiles; the latter is obtained from deep Chandra observations, as multiple dPIE fits, as described in \bo; 2) a clumpy component representing the cluster member galaxies (DM+baryons), modeled as spherical dPIE halos; 3) a shear+foreground-structure term to take into account the presence of massive structures in the outer cluster regions and line-of-sight mass distributions. Specific details on these multiple components are given below in the description of each lens model. The total cluster gravitational potential has therefore the form:

\begin{equation}
    \label{eq.: mass_decomposition}
    \phi_{tot} = \sum_{i=1}^{N_{h}} \phi_i^{halo}+ \sum_{j=1}^{N_{g}} \phi_j^{gal}+\phi_{shear+foreg},
\end{equation}
where the first sum runs over the $N_h$ cluster-scale mass distributions, while the second on the $N_g$ cluster member galaxies. \\

The general functional form for the spherical dPIE, including the relations between 3D and projected mass densities, as well as the expressions to derive aperture projected line-of-sight velocity dispersions, are given in Appendix\,\ref{sec.:appendix_dPIE_mass}. 
Following our previous models and the general {\texttt{LensTool}} methodology, the dPIE parameters for the sub-halo population follow a scaling relation for the central velocity dispersion and the truncation radius:

\begin{equation}
    \sigma^{gal}_{LT,i}= \sigma^{ref}_{LT} \left(  \frac{L_i}{L_0} \right)^{\alpha},
    \label{eq.: Scaling_relation_sigma}
    \end{equation}
    \begin{equation}
    r^{gal}_{cut,i}= r^{ref}_{cut} \left(  \frac{L_i}{L_0} \right)^{\beta_{cut}}, 
    \label{eq.: Scaling_relation_rcut}
\end{equation}

\noindent where $L_{i}$ is the luminosity of the i-th cluster member and $r^{gal}_{cut,i}$
represents the corresponding truncation radius. A similar scaling relation for the core radius, $r_{core}$, is used in \texttt{LensTool}, however it is not relevant here since a vanishing core radius is adopted. $\sigma^{gal}_{LT,i}$ is the {\texttt{LensTool}} fiducial velocity dispersion of each member, which is related to the central velocity dispersion of the dPIE profile by
$\sigma^{gal}_{0,i}= \sqrt{3/2}\, \sigma^{gal}_{LT,i}$ (see Appendix\,\ref{sec.:appendix_dPIE_mass}). 
It is common practice to fix the slopes $\alpha$ and $\beta_{cut},$ so that the model optimization is  performed  over only  two free parameters, that is the normalizations of the velocity dispersion and the truncation radius corresponding to the reference luminosity $L_0$. We measure the luminosities $L_i$ and $L_0$ using the HST F160W Kron magnitudes, which are a good proxy of the stellar mass of the cluster members (see \citealt{Grillo_2015}) and include members down to $m_{F160W}=24$. This leads to a minimum of 193 sub-halos to be included in the lens model (see Table\,\ref{table:cluster_sample}), fixed at the galaxy positions. 

The total mass of a circular dPIE profile is given by (see appendix\,\ref{sec.:appendix_dPIE_mass}, \citealt{Eliasdottir_lenstool}, \citealt{Limousin_lenstool}) the relation: $M_{tot} = \pi \sigma_{0}^2 r_{cut}/G $. 

\noindent Assuming a fixed scaling between the cluster members luminosity $L_i$ (in the same band considered by Eqs.\,\ref{eq.: Scaling_relation_sigma} and \ref{eq.: Scaling_relation_rcut}) and its total mass $M_{tot,i}$, that is $M_{tot,i}/L_i\propto L_i^\gamma$, one can obtain the following relation between the slopes of Eqs.\,\ref{eq.: Scaling_relation_sigma} and \ref{eq.: Scaling_relation_rcut}:

\begin{equation}
    \label{eq.: slopes}
    \beta_{cut}=\gamma-2\alpha+1.
\end{equation}

\begin{figure}[t!]
	\centering
	\includegraphics[width=9cm]{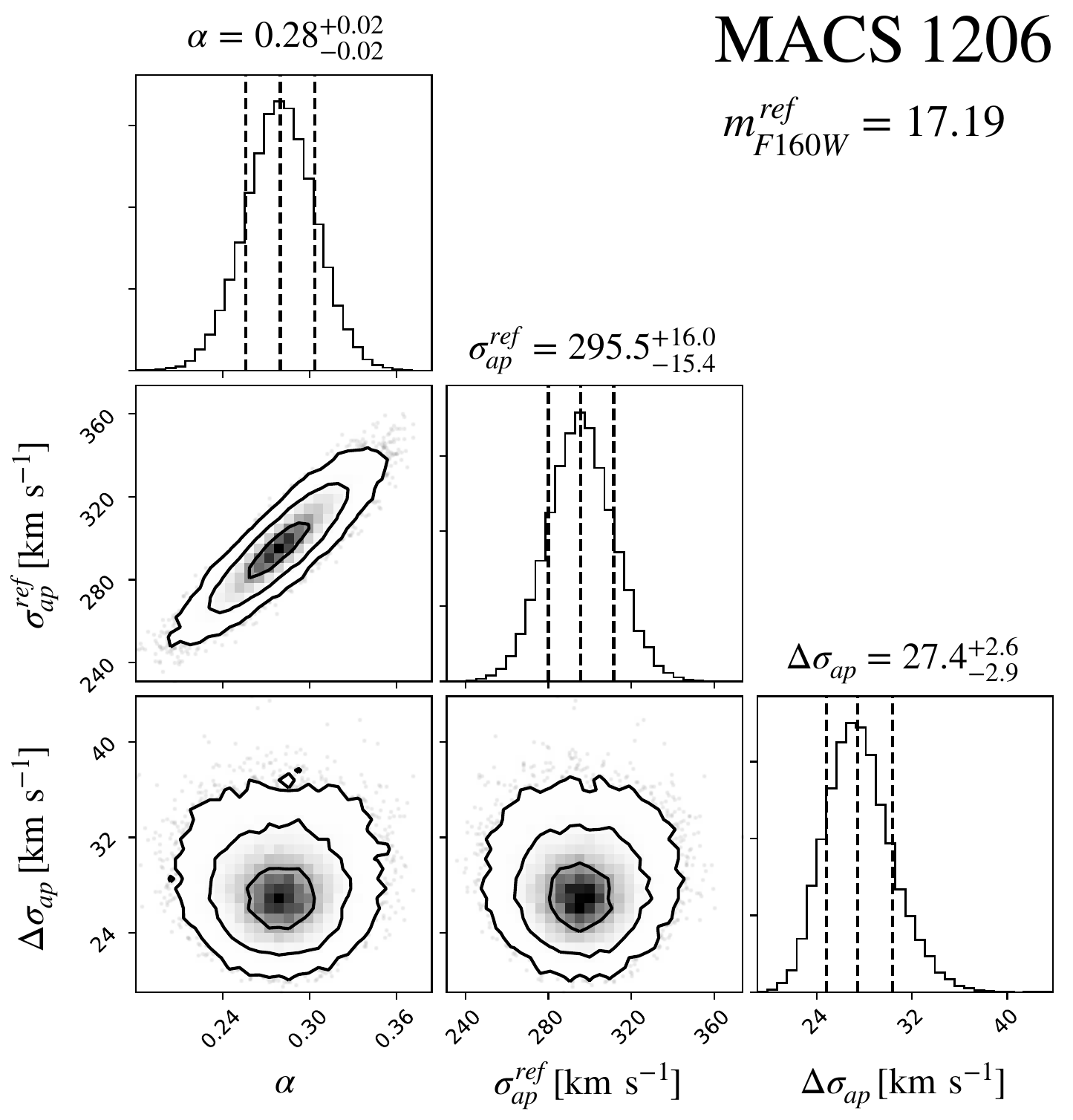}
	\caption{Posterior probability distributions for $\sigma_{ap}\mbox{-}mag$ scaling relation parameters, obtained as described in Appendix\,\ref{sec.:appendix_emcee}, from velocity dispersion measurements of 58 cluster members in \ma. The $16^{th}$, $50^{th}$ and $84^{th}$ percentiles of the marginalized distributions for the slope ($\alpha$), normalization ($\sigma_{ap}^{ref}$) and scatter around the scaling relation ($\Delta \sigma_{ap}$) are quoted and shown as vertical dashed lines.}
	\label{fig:emcee_macs1206}
\end{figure}

The main goal of this work is to use prior information for the scaling relations derived directly from measured velocity dispersions of cluster galaxies. The latter are light-weighted projected values of the 3D-velocity dispersions within the extracted spectroscopic apertures, hereafter $\sigma_{ap}$. In order to compare $\sigma_{LT}\,\mbox{-}\,L$ scaling relations with measured quantities, we need to compute the aperture-averaged line-of-sight velocity dispersion from the adopted dPIE mass models. The projection coefficients needed to transform $\sigma_{LT}$ into $\sigma_{ap}$ are obtained with a numerical integration depending on $r_{cut}$, $r_{core}$ and $R_{ap}$. In Appendix\,\ref{sec.:appendix_dPIE_mass} we recall all the equations to perform this projection procedure.

In Fig.\,\ref{fig:green_plots}, we show the best-fit scaling relations obtained from the previous models for each cluster (red and blue curves), which did not include any prior from internal kinematics of cluster members. These curves are computed by projecting the {\texttt{LensTool}} scaling relations, using the posterior distribution of the parameters $r_{cut}$ and $\sigma_{LT}^{ref}$ at different magnitudes, yielding the model $\sigma_{ap}$ as a function of $m_{F160W}$. In this process, we use an aperture of 0.8\arcsec\ so that the projected scaling relations can be directly compared with our kinematic measurements.

As customary in previous lens models to date, all the slopes for scaling relations $\alpha,\,\beta_{cut},\,\gamma$ (see Eq.\,\ref{eq.: slopes}) are fixed, here instead we take advantage of the measured stellar velocity dispersions to directly fit the normalization $\sigma_{ap}^{ref}$ and slope $\alpha$ of the $\sigma\,\mbox{-}\,L$ scaling relation. We then  derive $\beta_{cut}$ by  adopting $\gamma=0.2$, which is consistent with the canonical fundamental plane (\citealt{Faber_1987}, \citealt{Bender_1992}), and that we verify to be appropriate using photometric and morphological data for \ab\ (see below and Mercurio et al. in prep.). 

In order to fit the $\sigma\,\mbox{-}\,mag$ relation, a Bayesian approach is used, as described in Appendix\,\ref{sec.:appendix_emcee}. Thus, we derive the value of $\alpha$ and $\sigma_{ap}^{ref}$, as shown in Fig.\,\ref{fig:emcee_macs1206}, which also includes the intrinsic scatter $\Delta\sigma_{ap}$ of the measured velocity dispersions around the model (Eq.\,\ref{eq.: Scaling_relation_sigma}). Similar behaviors for the parameters derived for the other two clusters can be found in Appendix\,\ref{sec.:appendix_emcee}. The best-fit scaling relations obtained with this method are shown in Fig.\,\ref{fig:green_plots}, including the corresponding uncertainties (green curves), which are derived by sampling 300 times the posterior distributions. These best-fit parameters for each scaling relation (see Table\,\ref{table:kinemacs_scaling_relations}) are then used as dynamical priors in our lens models, as described below. 
It is interesting to note that the slope that we obtain for the Faber-Jackson relation, $L\,\mbox{-}\,\sigma^{1/\alpha}$, is very similar for the three clusters ($\alpha=0.27\,\mbox{-}\,0.28$) and consistent with several spectro-photometric studies of cluster early-type galaxy populations in the literature (e.g., \citealt{Kormendy_2013},\ \citealt{Focardi_2012}). 

We note that some galaxies deviate significantly from the best-fit scaling relations (as marked by boxes and circles in Fig.\,\ref{fig:green_plots}). This is however expected, as the Faber-Jackson relation is one of the projections of the fundamental plane relation among half-light radius $R_{e},$ mean surface brightness $\mu_e$ within $R_{e},$ and velocity dispersion: at a given luminosity, more compact galaxies tend to have higher velocity dispersion. 
For example, the four cluster galaxies at $m_{F160W}\sim 19$ in \ab, with a $\sigma$ significantly higher than the best-fit relation (see boxes in Fig.\,\ref{fig:green_plots}), have $\mu_e$ in the 16-th highest percentile, however we verified that they still lie on the fundamental plane defined by \cite{Jorgensen_96} (Mercurio et al., in prep.).

On the other hand, the few galaxies lying well below the $\sigma\,\mbox{-}\,mag$ relation (see circled data points in Fig.\,\ref{fig:green_plots} and the spectrum in the upper panel of Fig.\,\ref{fig:pPXF_spectra}) show sign of emission lines and young stellar populations in their spectra, for which lower velocity dispersions are expected when compared to early-type galaxies with similar luminosities.
 
\begin{table*}  
	\tiny
	\def\arraystretch{1.6}
	\centering    
	\begin{tabular}{|c|c|c|c|c|c|}
	   \hline
	   \textbf{Cluster} & \boldmath{$m_{F160W}^{ref}$} & \boldmath{$\sigma_{ap}^{ref}\ \mathrm{[km\ s^{-1}]}$} & \boldmath{$\alpha$} & \boldmath{$\Delta\sigma_{ap}\ \mathrm{[km\ s^{-1}]}$} & \boldmath{$\beta_{cut}(\gamma=0.2)$} \cr
	   \hline
	   \hline
	   \bf{MACS J1206.2$\mathbf{-}$0847} & 17.19 & $295.5_{-15.4}^{+16.0}$ & $0.28_{-0.02}^{+0.02}$  & $27.4_{-2.6}^{+2.9}$ & $0.64_{-0.04}^{+0.04}$ \cr
	   \hline
	   \bf{MACS J0416.1$\mathbf{-}$0403} & 17.02 & $281.2_{-16.0}^{+16.8}$ & $0.27_{-0.03}^{+0.03}$  & $31.4_{-3.1}^{+3.7}$ & $0.66_{-0.06}^{+0.06}$ \cr
	   \hline
	   \bf{Abell S1063} & 16.18 & $347.9_{-38.0}^{+39.5}$ & $0.27_{-0.04}^{+0.04}$  & $44.8_{-5.1}^{+6.1}$ & $0.66_{-0.08}^{+0.08}$ \cr
	   \hline
	\end{tabular}
	\smallskip
    \caption{$\sigma_{ap}\mbox{-}L$ scaling relation parameters derived from measured velocity dispersions of cluster members for \ma, \mb\ and \ab. The normalization parameter, $\sigma_{ap}^{ref}$, is computed at the reference magnitudes $m_{F160W}^{ref}$. Median values are derived from the marginalized parameter distributions, while the errors correspond to the $16^{th}$ and $84^{th}$ percentiles. The $\beta_{cut}$ values are obtained using Eq.\ref{eq.: slopes}}.    
	\label{table:kinemacs_scaling_relations} 

\end{table*}

\section{Combining lensing models with kinematics measurements}
\label{sec.:Lensing_kinematics}

In Fig.\,\ref{fig:green_plots}, a comparison between the sub-halo scaling relations obtained from the lens models and those directly constrained from kinematic measurements (green curves) shows significant discrepancies in the case of \mb\ and \ab, while they are consistent for \ma.
In the case of \ab, the normalization of the kinematic scaling relations are found to be approximately 100 and 150\,\vel\,above the values inferred from the \bo\ and \caab\ lens models, respectively. In \mb, the discrepancy is significant ($\sim$50\,\vel) for the \camb\ model, while it is negligible for the \bo\ model within the errors, albeit with a slope which deviates from the observed one ($\alpha =0.27$ against 0.35 assumed in \bo). This shows how inherent degeneracies of sub-halo population parameters in strong lensing models can lead to inferred velocity dispersion normalizations which are inconsistent with kinematic measurements of cluster galaxies. Nevertheless, these lens models can reproduce the positions of the multiple images with high precision, with a root-mean square value between the observed and model-predicted images on the image plane of $\Delta_{rms}\simeq 0.45\,\mbox{-}\,0.6\arcsec$ (see Table\,\ref{table:outputs_lensing}). It is in fact well known that parametric cluster lens models are, in general, affected by some degeneracy between the mass distribution of the macro-halo(s) and that of the sub-halos, even when a large number of constraints are available, as in our case \citep{Meneghetti_2017}. Despite that, the projected total mass value within a given cluster-centric radius remains robust. In addition, a significant degeneracy exists between the central velocity dispersions and the cut-off radii of the sub-halos. \\

In the following, we describe in detail our new lens models for each cluster. We start from the same parameterization and input constraints as in the \bo\ models, we then proceed to add critical  constraints on the sub-halo scaling relation parameters, with priors from our kinematic measurements. 
The optimization of model parameters is obtained from MCMC chains of approximately $10^5$ samples, excluding the burn-in phase. Model components and parameters are summarized in Table\,\ref{table:outputs_lensing}. As customary in the literature, no scatter in the scaling relations is assumed in our lens models, as it is currently not possible to include it in \texttt{LensTool} (Bergamini et al. in prep.) \\

\subsection{\ma}
Following \cama\ and \bo\ models, the cluster smooth mass distribution includes three dark matter halos which can reproduce the apparent elongated asymmetry in the distribution of the cluster galaxies and intra-cluster light. The three halos are described as dPIEs profiles whose values of sky positions, ellipticities, position angles, core radii, and velocity dispersions are left free to vary with flat priors, while their truncation radii are fixed to a large value. Three dPIE gas clumps are used to model the X-ray surface brightness distribution as in \bo. Thus, the cluster-scale mass components have a total of 18 free parameters. The presence of an external shear term introduces two extra free parameters in the model. Finally, the clumpy component includes 258 halos describing the cluster members and the BCG, centered on the peaks of their light emission. All these galaxies are described as circular dPIEs profiles, whose values of central velocity dispersion and truncation radius scale with their F160W magnitude, according to Eqs.\,\ref{eq.: Scaling_relation_sigma} and \ref{eq.: Scaling_relation_rcut}, where the BCG magnitude is used as reference luminosity ($L_0$). The normalization $r_{cut}^{ref}$ is free to vary between 1\arcsec and 50\arcsec\ (i.e. $5.67\,\mbox{-}\,283.9$\,kpc at the cluster redshift) with a flat prior. A Gaussian prior derived from our kinematic measurements is then introduced for the $\sigma\,\mbox{-}\,L$ scaling relation. To determine this prior, we deproject the best-fit (median) normalization of the $\sigma\,\mbox{-}\,mag$ relation ($\sigma_{ap}^{ref}=295.5^{+16.0}_{-15.4}$\,\vel, see Fig.\,\ref{fig:emcee_macs1206}) to obtain a \texttt{LensTool} fiducial reference velocity dispersion, $\sigma_{LT}^{ref}$. This deprojection is achieved with an iterative procedure by computing the projection coefficient, $c_p(r_{core}, r_{cut}, R_{ap})$ (see Appendix~\ref{sec.:appendix_dPIE_mass}), using the best-fit scaling relation parameters from a first model without kinematic prior, as $\sigma_{LT}^{ref}=\sigma_{ap}^{ref}/c_p$.
In the case of \ma, we thus obtain a Gaussian prior on $\sigma_{LT}^{ref}$ with a mean of $264$\,\vel\, and a standard deviation of $18$\,\vel. The standard deviation also includes the uncertainty in the computation of $c_p$, due to the allowed range of $r_{cut}$. Finally, the slope $\alpha$ is fixed to the value obtained from the $\sigma\,\mbox{-}\,mag$ fit ($\alpha=0.28$), while a $\beta_{cut}=0.64$ is obtained  from Eq.\,\ref{eq.: slopes}, with $\gamma= 0.2$.
The lens model is thus optimized with a total number of 22 free parameters, using the observed positions of 82 spectroscopically confirmed multiple images associated to 27 families (C17b and \bo).\\

\subsection{\mb}
Referring to Table\,\ref{table:outputs_lensing}, the cluster-scale mass components of this merging cluster (\citealt{Balestra_MACS0416}) include two massive dark matter halos, whose positions are left free to vary in areas of $30\arcsec\times 30\arcsec$ and $15\arcsec\times 15\arcsec$ around the northern and southern BCG, respectively (BCG,N, BCG,S); plus a third circular halo in the NW region to improve the model accuracy ($\Delta_{rms}$) in that area. Four dPIEs with fixed parameters are used to describe the complex X-ray emitting gas distribution in \mb, as done in \bo. The cluster-scale component has therefore 16 free parameters. No shear is present, however an additional circular dPIE halo with free velocity dispersion and truncation radius is added to account for the presence of a foreground galaxy (\citealt{Caminha_macs0416}, \citealt{Jullo_lenstool}).\\
The sub-halo mass component, including the BCG, comprises 193 halos in this case. The $r_{cut}^{ref}$ parameter is left free to vary over a wide range (1\,-\,20\arcsec, or 5.3\,-\,106.8\,kpc). As described above for \ma, the deprojection of the best-fit $\sigma_{ap}\,\mbox{-}\,mag$ relation provides a Gaussian prior for the normalization of $\sigma_{LT}^{ref}=249\pm 15$\,\vel, while we use the measured slope $\alpha=0.27$ (see Fig.\,\ref{fig:emcee_two}/top), from which we derive $\beta_{cut}=0.66$ (from Eq.\,\ref{eq.: slopes}, with $\gamma=0.2$).
With a total number of 20 free parameters, 
102 spectroscopically confirmed multiple images (associated to 37 families) are then used to constrain the lens model (as in C17a and \bo). \\

\subsection{\ab}
The dark-matter macro-halo includes in this case an elliptical dPIE, close to the BCG position, and a circular dPIE whose position is left free to vary in an area of $150\arcsec \times 120\arcsec$ centered in the north-east region of the cluster (Table\,\ref{table:outputs_lensing}). Three dPIEs are used to describe the gas component (\bo), and no shear term is included. There are therefore nine free parameters for the cluster-scale component. \\
The clumpy sub-halo component is constituted by 222 cluster members, including the BCG. From our kinematic measurements and best-fit scaling relation we obtain a deprojected value of $\sigma^{ref}_{LT} = 310_{-34}^{+35}$\,\vel. As a result, we adopt the median value of $310$\,\vel\, as a Gaussian prior for the normalization of the $\sigma_{ap}\,\mbox{-}\,L$ scaling relation, while the standard deviation is reduced to $15$\,\vel\, to force the lens model to converge to a $\chi^2$ minimum solution compatible with our galaxy kinematics. The slope is fixed to the measured value $\alpha=0.27$ (see Fig.\,\ref{fig:emcee_two}/bottom), which implies $\beta_{cut}=0.66$.
The lens model, which has 11 free parameters in total, is thus optimized using the positions of 55 spectroscopically confirmed multiple images, divided in 20 families, as implemented in \bo.\\

\section{Results and discussion}
\label{Results}
\begin{figure*}[ht!]
	\centering
	\includegraphics[width=\linewidth]{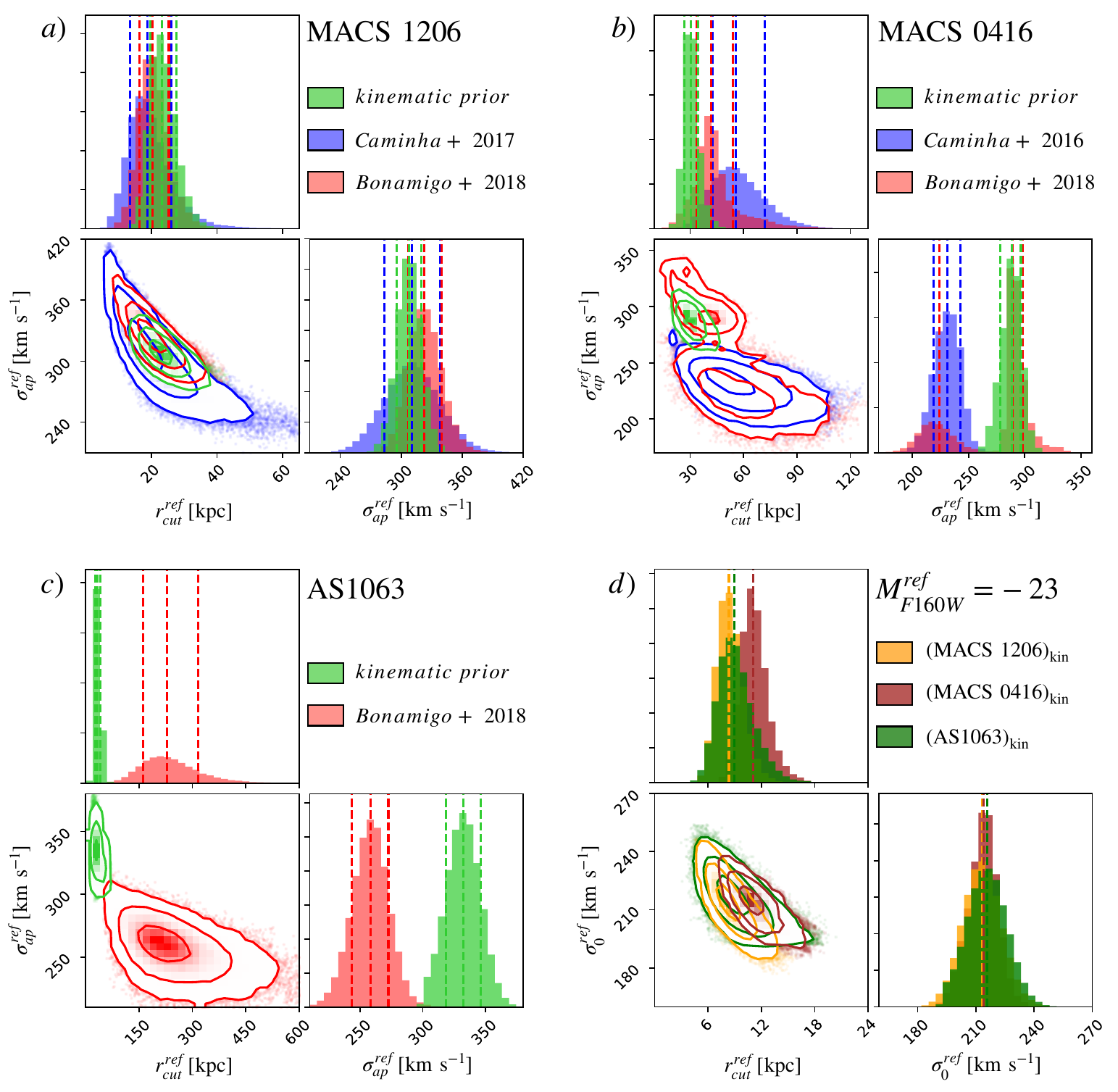}
	\caption{Panels a,b,c:  posterior probability distributions for the normalization of the $\sigma_{ap}\mbox{-}L$ scaling relation, that is aperture projected velocity dispersion $\sigma_{ap}^{ref}$, and truncation radius ($r^{ref}_{cut}$), obtained with different lens models for clusters under study. The green contours and distributions refer to our best models which include the kinematic prior based on measured velocity dispersions of member galaxies (Fig.\,\ref{fig:green_plots}). For visualization clarity, in panel c we omit the \caab\ results because they are outside the chosen range of velocity dispersion, as visible in the bottom panel of Fig.\ref{fig:green_plots}. The reference magnitudes for each cluster are in Table\,\ref{table:kinemacs_scaling_relations}. In panel (d), we show the posterior distribution of the reference central velocity ($\sigma_{0}^{ref}$) and reference truncation radius ($r^{ref}_{cut}$) in each cluster, normalized to an absolute mag $M^{ref}_{F160W}=-23$, close to $L^\ast$ of the early-type galaxy population. The different colors correspond to the models with kinematic priors and contours refer to 1, 2, 3$-\sigma$ confidence levels. Once normalized to the same luminosity, the three scaling relations for $\sigma$ and $r_{cut}$ are consistent. }
	\label{fig:degeneracies_all}
\end{figure*}

The inclusion of the kinematic prior in the lens models has the main consequence of significantly reducing the intrinsic degeneracies in the scaling relation parameters. Specifically, since the projected mass of the sub-halos depends on the central velocity dispersion, $\sigma_0$, and truncation radius $r_{cut}$, for a vanishing core radius, there is clearly an inherent degeneracy between the normalization parameters of the $\sigma\,\mbox{-}\,L$ and $r_{cut}\,\mbox{-}\,L$ scaling relation. This can be easily appreciated from the expression of the dPIE projected mass within a radius $R$, for $r_{core}=0$ (Eq.\,\ref{eq.: Mass_2d_dPIE} in Appendix\,\ref{sec.:appendix_dPIE_mass}, or \citealt{Jullo_lenstool}):
\begin{equation}
    M(R)=\frac {\pi \sigma_0^2}{G}\left(R+r_{cut}-\sqrt{r_{cut}^2+R^2}\right).
\end{equation}
Therefore, several combinations of $\sigma_0$ and $r_{cut}$ will yield similar aperture masses, which are constrained by the multiple image positions. This degeneracy, in combination with those among the parameters describing cluster-scale mass components or shear terms, can lead the lens model to predict a distribution of sub-halo masses not consistent with internal kinematics of member galaxies. 
This can be noticed in Fig.~\ref{fig:green_plots} where we display the case of \mb\ and \ab. The implementation of the kinematic prior on the normalization and slopes of the scaling relation leads to models with a global precision which is similar to that of the previous lens models, as apparent from a comparison of $\Delta_{rms}$ values (see Table\,\ref{table:outputs_lensing}). The new values compared to the \bo\ models differ by 0.01\arcsec\ for \ma, 0.02\arcsec\ for \mb, and 0.04\arcsec\ for \ab, with the same number of degrees of freedom (88 for \ma, 110 for \mb\ and 59 for \ab). The slight increase of $\Delta_{rms}$ in the new models is expected since the freedom of model parameters is generally reduced with the Gaussian priors.

In Fig.\,\ref{fig:degeneracies_all}, we show the posterior probability distributions for the projected values of the normalization of the $\sigma\,\mbox{-}\,L$ and $r_{cut}\,\mbox{-}\,L$ relations, $\sigma_{ap}^{ref}$ and $r_{cut}^{ref}$, respectively, for the new (in green) and previous models. The kinematic priors lead in general to significant smaller uncertainties for these parameters, up to factor of 10 for the truncation radius of \ab, with respect to the \bo\ results.

\begin{figure}[hb!]
	\centering
	\includegraphics[width=9cm]{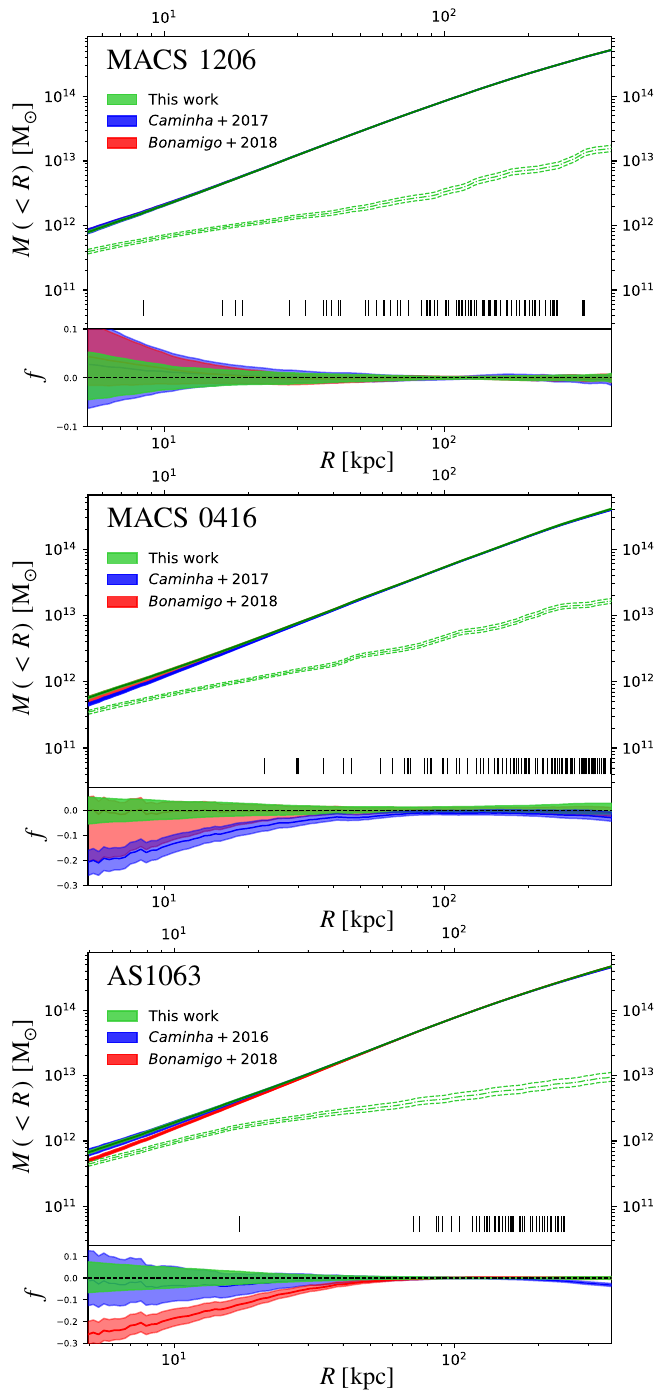}
	\caption{
    \textit{Top:} Projected cumulative mass profiles, as a function of the projected distance from BCGs, corresponding to previous and current lens models for each cluster. The colored regions encompass the $16^{th}$ and the $84^{th}$ percentiles; colored solid lines are the median values.
    The dashed green lines correspond to the mass component associated to cluster members (i.e. sub-halos) from the new lens models with kinematic prior ($16^{th}$, $50^{th}$, $84^{th}$ percentiles). The multiple image projected distances from the cluster centers are marked with vertical black lines. \textit{Bottom:} Relative variation of the cumulative projected total masses with respect to our reference (green) model. 
	}
	\label{fig:mass_profiles}
\end{figure}

 While for \ma\ the new solution is consistent with the kinematic data, for the other two clusters the kinematic prior moves the best-fit solution to a different region of the parameter space. In the case of \mb, the \bo\ model had already found a different solution with a higher $\sigma$-normalization with respect to the C17a, by extending the MCMC parameter search and including the hot gas mass component. Our new lens model is consistent within $1 \sigma$ with \bo, further reducing parameters' uncertainties especially on $r_{cut}$. In the case of \ab, the kinematic prior moves the $\sigma_{ap}^{ref}$ (for the BCG) to a much larger value, which is 70\,\vel\ higher, and provides a much tighter constraint for $r_{cut}^{ref}$, moving the solution away from previous values of $\sim\! 190$ kpc (a dPIE encircles 90\% of the total projected mass within $5 r_{cut}$). 

It is relevant to note that, despite the higher normalization of the sub-halo scaling relations for \mb\ and \ab, the cluster total projected mass profile does not vary appreciably when compared to previous models. This is expected since the multiple image positions provide information about the cluster total mass projected within circles with radii equal to the average distance of the multiple images of each family. In Fig.\,\ref{fig:mass_profiles}, one can appreciate that the differences in the cumulative projected mass between the new and previous models are well within 10\% over the radial region with multiple image constraints. 
Therefore, although the fraction of total mass in sub-halos differs for \mb\ and \ab, the total encircled mass remains a robust measurement. This however shows that the inclusion of the kinematic prior is important to obtain a reliable estimate of the relative contribution of the diffuse and clumpy (sub-halo) mass components from strong lensing models. In general, an underestimate  of the sub-halo mass component will lead to a higher contribution of the cluster-scale halos, in particular the central one. 

\subsection{Sub-halo scaling relations}

The consistency of the sub-halo scaling relations, based on kinematic measurements in the three clusters, suggests a rescaling of the normalization of the $\sigma\,\mbox{-}\,L$ and $r_{cut}\,\mbox{-}\,L$ relations to the same absolute luminosity for a meaningful comparison. To this aim, we choose as reference an absolute magnitude of $M^{ref}_{F160W}=-23$, which is close to the value of $L^\ast$, estimated from the F160W luminosity function of \mb\ at z=0.4 (\citealt{Connor_2017} and Mercurio et al. in prep.). We then rescale the normalization of the scaling relations obtained from the kinematic lens models, anchored on the BCG magnitude ($m^{ref}_{F160W}$), to the new reference $M^{ref}_{F160W}$ using the distance modulus $\mu$ of each cluster, as follows: 
\begin{equation}
    \sigma_{0}^{M} = \sigma_{0}^{ref} 10^{0.4\, [m^{ref}_{F160W}-\mu(z_{cl})-M^{ref}_{F160W}]\, \alpha},
\end{equation}
\noindent where the central reference velocity dispersion is  $\sigma_{0}^{ref}=\sqrt{3/2}\, \sigma_{LT}^{ref}$ in Eq.\,\ref{eq.: Scaling_relation_sigma}, and

\begin{equation}
    r_{cut}^{M} = r_{cut}^{ref} 10^{0.4\, [m^{ref}_{F160W}-\mu(z_{cl})-M^{ref}_{F160W}]\, \beta_{cut}}.
\end{equation}

The comparison of the posterior distributions of $\sigma_0,\,{\rm and}\, r_{cut}$, obtained from our lens models with kinematic prior, once renormalized to the same $L^\ast$ luminosity (i.e. $M^{ref}_{F160W}$), is shown in the bottom right panel (d) in Fig.\,\ref{fig:degeneracies_all}. The consistency of these distributions among the three clusters is quite remarkable, particularly for $r_{cut}$, and suggests an empirical relation for the truncation radius, which is generally poorly constrained by lens models without kinematic priors. By combining Eqs.~\ref{eq.: Scaling_relation_sigma} and \ref{eq.: Scaling_relation_rcut}, one obtains: 
\begin{equation}
    r_{cut} = 10.1_{(7.3)}^{(13.1)}\,{\rm kpc}\,\left(\frac{\sigma_0}{220 \,({\rm km\,s^{-1}})}\right)^{2.43_{(2.38)}^{(2.45)}},
    \label{eq. rcut-sigma}
\end{equation}
where the range for each parameter represents the $16^{th}\mbox{-}84^{th}$ percentiles of the combined posterior distributions of $\sigma_0^{ref},\,{\rm and}\, r_{cut}^{ref}$, and the slope is obtained by the distribution of $\beta_{cut}/\alpha$. 
A similar relation was derived by \cite{Monna_2015} for Abell 383 ($z=0.18$) incorporating the velocity dispersion measurements of 21 member galaxies. However, in that study the slopes of the $\sigma\propto L^\alpha$ and $r_{cut}\propto L^{\beta_{cut}}$ relations are fixed, based on literature studies, assuming $M_{tot}/L=const$. In our case, the slope $\alpha$ is directly measured from the MUSE spectra of a cluster galaxy sample larger by more than a factor of 2, while $\beta_{cut}$ is derived assuming galaxies lying on the fundamental plane. 

The slope $\beta_{cut}/\alpha=2.43$ that we find for the $r_{cut}\,\mbox{-}\,\sigma_0$ relation is likewise the result of our kinematic measurements in three different clusters, whereas the slope of 1.25 quoted in \cite{Monna_2015} is derived from the assumed values of $\alpha$ and $\beta_{cut}$. However, the normalization of the relation in Eq.\,\ref{eq. rcut-sigma} turns out to be consistent with the \cite{Monna_2015} result. 
Other attempts to constrain halo sizes of cluster galaxies have involved  modeling of single strong lensing systems in clusters, with no kinematics information from spectroscopic measurements. For example, \cite{Suyu_2010} found a $r_{cut}=6.0_{-2.0}^{+2.9}$\,kpc for a galaxy with a lens-based velocity dispersion $\sigma_0=127_{-12}^{+21}$\,\vel\ in a group, in agreement with our results. \cite{Eichner_2013} found instead a higher normalization by modeling the surface brightness distribution of the "snake arc" in \ma. However, we should note that the lens model used in that work was based on a first limited sample of multiple images, when MUSE spectroscopy was not available. \cite{Limousin_2007_truncation} estimated a scaling relation between halo sizes and $\sigma_0$ using galaxy-galaxy weak lensing in 5 clusters at $z\sim 0.2$, which is broadly consistent with our results.

The consistent constraints we find on the central velocity dispersions and truncation radii of the cluster sub-halos from the independent analysis of three clusters suggest that we can combine Eqs.\,\ref{eq.: Total_mass_dPIE} and \ref{eq. rcut-sigma} to obtain an empirical relation between $\sigma_0$ and the total mass of member galaxies. By propagating the uncertainties derived from the posterior distributions of  $\sigma_0$ and $r_{cut}$,  we obtain the following $M_{tot}\,\mbox{-}\,\sigma_0$ relation:
\begin{equation}
    M_{tot}=3.5_{(2.6)}^{(4.6)} \times 10^{11}\,\rm M_{\odot}\,\left(\frac{\sigma_0}{220\,\rm(km\,s^{-1})}\right)^{4.43^{(4.45)}_{(4.38)}},
    \label{eq. mass-sigma}
\end{equation}
\noindent where the parameter range refers to the $16^{th}$ and $84^{th}$ percentiles. 

\begin{table*}[h!]     
	\tiny
	\def\arraystretch{1.6}
	\centering          
	\begin{tabular}{|c|c|c|c|c|c|c|c|c|c|c|}
	    \cline{4-10}
		\multicolumn{3}{c|}{} & \multicolumn{7}{c|}{ \textbf{dPIE parameters}} \\
		\cline{4-11}
		  \multicolumn{3}{c|}{} & \boldmath{$x\, \mathrm{[arcsec]}$} & \boldmath{$y\, \mathrm{[arcsec]}$} & \boldmath{$e$} & \boldmath{$\theta\ [^{\circ}]$} & \boldmath{$\sigma_{LT}\, \mathrm{[km\ s^{-1}]}$} & \boldmath{$r_{core}\, \mathrm{[arcsec]}$} & \boldmath{$r_{cut}\, \mathrm{[arcsec]}$} & \boldmath{$\Delta_{rms}\, \mathrm{[arcsec]}$}\\ 
          \hline
		  \multirow{8}{*}{\rotatebox[origin=c]{90}{\textbf{MACS J1206.2$\mathbf{-}$0847}}} & \multirow{6}{*}{\rotatebox[origin=c]{90}{\textbf{HALOS}}} & $1^{st}$DM & $-0.75_{-0.43}^{+0.40}$ & $0.35_{-0.21}^{+0.20}$ & $0.69_{-0.02}^{+0.02}$ & $19.92_{-0.80}^{+0.84}$ & $786.9_{-35.0}^{+30.6}$ & $6.51_{-0.48}^{+0.45}$ & 2000.0 & \multirow{8}{*}{\shortstack{\textbf{This\ work:} 0.46 \\\\ \textbf{\bo:} 0.45 \\\\ \textbf{\cama:} 0.44}} \\
		  & & $2^{nd}$DM & $9.68_{-0.73}^{+0.78}$ & $3.88_{-0.72}^{+0.74}$ & $0.56_{-0.10}^{+0.09}$ & $114.65_{-2.38}^{+2.75}$ & $620.9_{-26.5}^{+27.3}$ & $14.37_{-1.18}^{+1.50}$ & 2000.0 &  \\
		  & & $3^{rd}$DM & $-28.46_{-1.43}^{+1.38}$ & $-6.61_{-0.74}^{+0.78}$ & $0.34_{-0.06}^{+0.06}$ & $-23.83_{-12.07}^{+10.59}$ & $491.4_{-31.9}^{+37.4}$ & $12.02_{-1.91}^{+2.15}$ & 2000.0 &  \\
		  & & $1^{st}$Gas & 3.11 & $-$6.34 & 0.12 & $-$0.71 & 452.2 & 63.29 & 403.05 &  \\
		  & & $2^{nd}$Gas & $-$13.50 & $-$7.24 & 0.50 & $-$113.57 & 342.3 & 40.53 & 43.94 &  \\
		  & & $3^{rd}$Gas & 3.31 & 2.04 & 0.58 & $-$169.20 & 186.9 & 8.24 & 68.57 &  \\
		  \cline{2-10}
		  & \rotatebox[origin=c]{90}{\textbf{S}} & Shear & - & - & $0.12_{-0.01}^{+0.01}$* & $101.15_{-1.36}^{+1.41}$ & - & - & - &  \\
		  \cline{2-10}
		  & \rotatebox[origin=c]{90}{\textbf{G}} & $258\, [17.19]$ & - & - & 0.0 & 0.0 & $272.6_{-12.6}^{+12.7}$ & 0.01 & $4.10_{-0.64}^{+0.76}$ &  \\
		  
		  \hline
		  \hline
		  
		  \multirow{8}{*}{\rotatebox[origin=c]{90}{\textbf{MACS J0416.1$\mathbf{-}$0403}}} & \multirow{7}{*}{\rotatebox[origin=c]{90}{\textbf{HALOS}}} & $1^{st}$ DM & $-2.14_{-0.84}^{+1.06}$ & $1.36_{-0.73}^{+0.63}$ & $0.84_{-0.04}^{+0.01}$ & $144.75_{-1.01}^{+1.12}$ & $581.0_{-24.0}^{+19.9}$ & $6.56_{-0.60}^{+0.59}$ & 2000.0 & \multirow{9}{*}{\shortstack{\textbf{This\ work:} 0.61 \\\\ \textbf{\bo:} 0.59 \\\\ \textbf{\camb:} 0.59}} \\
		  & & $2^{nd}$DM & $20.01_{-0.23}^{+0.22}$ & $-37.20_{-0.45}^{+0.44}$ & $0.76_{-0.01}^{+0.01}$ & $125.95_{-0.39}^{+0.46}$ & $859.9_{-15.4}^{+15.0}$ & $11.98_{-0.58}^{+0.60}$ & 2000.0 &  \\
		  & & $3^{rd}$DM & $-33.99_{-1.12}^{+0.93}$ & $8.38_{-0.85}^{+2.83}$ & 0.0 & 0.0 & $314.2_{-50.0}^{+47.7}$ & $5.58_{-2.76}^{+2.62}$ & 2000.0 &  \\
		  & & $1^{st}$Gas & $-$18.14 & $-$12.13 & 0.12 & $-$156.76 & 433.0 & 149.21 & 149.82 &  \\
		  & & $2^{nd}$Gas & 30.79 & $-$48.67 & 0.42 & $-$71.50 & 249.0 & 34.77 & 165.77 &  \\
		  & & $3^{rd}$Gas & $-$2.37 & $-$1.26 & 0.42 & $-$54.74 & 101.7 & 8.28 & 37.59 &  \\
		  & & $4^{th}$Gas & $-$20.13 & 14.74 & 0.40 & $-$49.32 & 281.8 & 51.67 & 52.34 &  \\
		  \cline{2-10}
		  & \rotatebox[origin=c]{90}{\textbf{S}} & Foreground & 31.96 & $-$65.55 & 0.0 & 0.0 & $178.0_{-15.0}^{+14.6}$ & 0.05 & $61.9_{-21.55}^{+25.07}$ & \\

		  \cline{2-10}
		  & \multirow{1}{*}{\rotatebox[origin=c]{90}{\textbf{G}}} & $193\, [17.02]$ & - & - & 0.0 & 0.0 & $262.0_{-10.2}^{+8.5}$ & 0.05 & $5.68_{-0.69}^{+0.81}$ &  \\

		  \hline
		  \hline

		  \multirow{6}{*}{\rotatebox[origin=c]{90}{\textbf{Abell S1063}}} & \multirow{5}{*}{\rotatebox[origin=c]{90}{\textbf{HALOS}}} & $1^{st}$DM & $1.40_{-0.23}^{+0.23}$ & $-0.74_{-0.17}^{+0.16}$ & $0.63_{-0.01}^{+0.01}$ & $-38.95_{-0.23}^{+0.22}$ &
		  $1162.4_{-6.7}^{+6.4}$ & $18.06_{-0.52}^{+0.53}$ &  2000.0 & \multirow{6}{*}{\shortstack{\textbf{This\ work:} 0.55 \\\\ \textbf{\bo}: 0.51 \\\\ \textbf{\caab:} 0.51}} \\
		  & & $2^{nd}$DM & $-50.16_{-4.41}^{+3.70}$ & $26.80_{-2.17}^{+2.36}$ & 0.0 & 0.0 & $221.4_{-22.3}^{+24.2}$ & 0.05 & 2000.0 &  \\
		  & & $1^{st}$Gas & 18.90 & $-$73.36 & 0.80 & $-$162.05 & 335.9 & 188.40 & 189.24 &  \\
		  & & $2^{st}$Gas & $-$18.05 & 13.47 & 0.13 & $-$27.80 & 442.6 & 36.32 & 339.16 &  \\
		  & & $3^{rd}$Gas & 0.20 & $-$1.24 & 0.34 & $-$15.49 & 249.7 & 14.43 & 356.50 &  \\
		  \cline{2-10}
		  & \multirow{1}{*}{\rotatebox[origin=c]{90}{\textbf{G}}} 
		  & $222\, [16.18]$ & - & - & 0.0 & 0.0 & $299.4_{-14.2}^{+14.3}$ & 0.05 & $6.83_{-1.32}^{+1.69}$ &  \\
		  \hline

	\end{tabular}
	\smallskip
    \caption{Output parameters of three lens models developed in this work. The mass components are grouped into (\textbf{HALOS}), shears and foreground galaxies (\textbf{S}), and cluster galaxies following the scaling relations (\textbf{G}). Parameters with no errors are fixed. All cluster-scale halos are modeled as dPIEs. Halos making up the hot gas component are taken from \bo.  Sky $x,y$ coordinates are the offsets in arcsec from the reference BCG positions (\ma: $R.A.=12^h06^m12^s\!.15$, $DEC= -8^{\circ}48'03''\!.4$, \mb: $R.A.=04^h16^m09^s\!.15$, $DEC= -24^{\circ}04'02''\!.9$ and \ab: $R.A.=22^h48^m43^s\!.97$, $DEC= 44^{\circ}31'51''\!.2$). The ellipticity $e$ is defined as $e=\frac{a^2-b^2}{a^2+b^2}$, where $a$ and $b$ are the semi-major and semi-minor axis; when a shear term is present the ellipticity refers to $\gamma_{shear}$ (see \cama). The position angle $\theta$ is computed from west to north; $\sigma_{LT}$ is the {\texttt{LensTool}} fiducial velocity dispersion, $r_{core}$ and $r_{cut}$ are the core and truncation radius, respectively. For the galaxy component (G), the number of spherical sub-halos included in the model and the normalization F160W magnitude are quoted in the third column. For each optimized parameter, we quote the median, and the $16^{th}$ and $84^{th}$ percentiles as errors.
    The last column shows the total root-mean-square separation between the model-predicted and observed positions of the multiple images.}    

	\label{table:outputs_lensing}

\end{table*}

\subsection{Sub-halo velocity dispersion function}
In  light of our results, we present a central velocity dispersion function of cluster members, which can be derived from the sub-halo component of the lens models. This was first presented, in the form of circular velocity function ($v_c=\sqrt{2}\sigma_0$, see Appendix\,\ref{sec.:appendix_dPIE_mass}) in \cite{Grillo_2015} for \mb\ and later extended by \bo\ to the same sample of three clusters of the present study, including the hot-gas components. The new sub-halo velocity functions presented here are particularly relevant since the new lens models incorporate a prior on the $\sigma\,\mbox{-}\,L$ scaling relation based on measured velocity dispersions of a large number of cluster members. 

\begin{figure}[ht!]
	\centering
	\includegraphics[width=9cm]{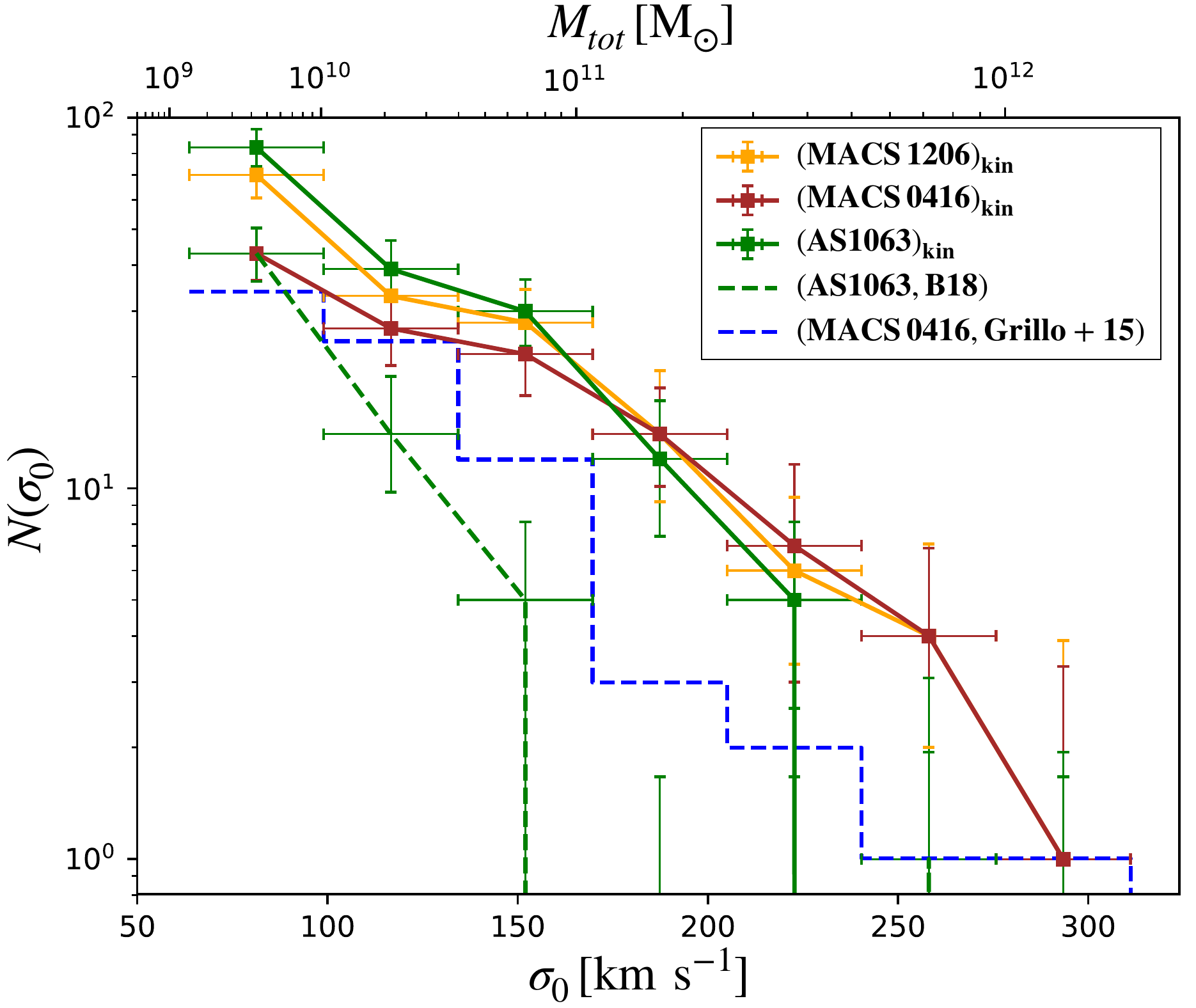}
	\caption{Sub-halo central velocity dispersion function derived from our lens models for \ma, \mb\ and \ab\ with kinematic prior based on measured velocity dispersions of cluster members. Previous models with no kinematic prior for \ab\ and \mb\ are shown for comparison. Data points correspond to median galaxy counts in 35\,\vel\, bins, drawn from the posterior distributions of sub-halo model parameters. Errors also include Poissonian number statistics in each bin. Only galaxies within a cluster-centric radius of $0.16 \times R_{200c}$ are included (see Table\,\ref{table:cluster_sample}). The top axis gives the total mass related to the central velocity dispersion by Eq.\,\ref{eq. mass-sigma} (see text).}
	\label{fig:velocity_distribution}
\end{figure}

Each data point in Fig.\,\ref{fig:velocity_distribution} is obtained by extracting the central velocity dispersion of each member from 500 random realizations of the model posterior distributions, thereby computing the median number of members in each velocity bin 35\,\vel\,wide. We adopt the same bin-size as in \bo\ to allow a direct comparison with previous results. The vertical error bars are obtained by combining the error associated to the $16^{th}$ and $84^{th}$ percentiles of these realizations with a Poissonian error generally appropriate for low number counts\footnote{$\Delta N^{up}=\sqrt{n+1}+2/3$ and $\Delta N^{down}=\sqrt{n}$, where $n$ is the number of objects per bin \citep{poisson}.}. In order to compare the velocity functions of the three clusters, we include only member galaxies with cluster-centric distances $R<0.16 \times R_{200c}$ (see Table \ref{table:cluster_sample}), which is the maximum aperture within which we have a highly complete sample of member galaxies in all three clusters.

In Fig.\,\ref{fig:velocity_distribution}, we compare the new velocity functions with the previous determinations by \cite{Grillo_2015} (for \mb, dashed blue line) and \bo\ (for \ab, dashed green line). Not surprisingly, these previous determinations were biased low due to the lower normalization of the scaling relations in previous models, whereas the new lens models with kinematic priors produce velocity functions which are quite consistent with each other.  

\section{Conclusions}
\label{Conclusions}
In this work, we have improved our previous strong lensing models of three Hubble Frontier Fields/CLASH clusters, based on a large number of spectroscopically confirmed multiple images, using robust measurements of the internal stellar velocity dispersion of large samples of clusters galaxies, thanks to MUSE integral-field spectroscopy in cluster cores. Such measurements constrain independently the normalization and slope of the Faber-Jackson $\sigma\,\mbox{-}\,L$ relation so that the sub-halo component of the cluster mass distribution in the lens models is now bound to the kinematic measurements, thus allowing a significant improvement on the reconstruction of the cluster internal sub-structure. 
We summarize below the main results of our study.

Using spectra of cluster early-type galaxies with a mean S/N of 25, we measure robust velocity dispersions of 37, 49 and 58 members in \ab, \mb, \ma, whose accuracy is tested with extensive spectral simulations. By obtaining kinematics measurements over 4\,-\,5 magnitudes in each cluster, down to $m_{F160W}\simeq 21.5$ which corresponds to $\sim\! 2.5$ mag below $L^\ast$, we sample well the Faber-Jackson relation. A maximum likelihood modeling of this relation yields well consistent normalizations and slopes for the three clusters. 

The new lens models incorporating kinematics information of cluster galaxies reproduce the positions of multiple images with similar precision to previous models. While the total projected cluster mass profile remains essentially unchanged, the mass of sub-halos are now robustly constrained, thus reducing degeneracies with other mass components and parameters of the lens model. 

Specifically, the inherent degeneracy of lens models between the central velocity dispersion ($\sigma_0$) and truncation radius ($r_{cut}$) of sub-halos is strongly reduced, thus providing robust estimates of sub-halo masses and sizes.  Once normalized to the same absolute luminosity, the three $\sigma_0\,\mbox{-}\,r_{cut}$ scaling relations, independently derived for each cluster, are found in very good agreement. This is particularly interesting in consideration of the different dynamical states of the three systems. As a result, we obtain a statistical determination of sub-halo truncation radii with masses ranging from that of the BCG down to $\sim\! 10^{10}\, M_\odot$. Our findings extend the initial results derived by \cite{Monna_2015} in A383 by using now a high-quality data set which includes precision lens models based on large number bona-fide multiple images and high S/N kinematic measurements of cluster members.   
    
Interestingly, with such a robust determination of the scaling relations of the sub-halo populations, we infer fully consistent velocity dispersion functions for the three clusters, unlike in previous lens models. In addition, our new constraints on the sub-halo masses provide an empirical $M_{tot}\,\mbox{-}\,\sigma_0$ relation which can be used to translate the velocity functions into sub-halo mass functions.

We plan to extend this methodology to a larger sample of CLASH clusters with strong lensing models based on MUSE observations \citep{Caminha_2019} to better explore the role of systematics in parametric lens models when constraining different mass components of galaxy clusters. Further constraints on the mass profile of cluster galaxies, including the dark matter fraction, will require the use of galaxy-scale strong lensing systems (e.g., \citealt{Suyu_2010}, \citealt{Grillo_2014}, \citealt{Monna_2015}), in combination with the internal kinematics of the lenses to extend in dense environments the extensive work carried out in field early-type galaxies. A complementary analysis based on the statistics of galaxy-galaxy strong lensing events will also be presented in an upcoming paper (Meneghetti et al. in prep.). We defer to future papers a detailed comparison of the sub-halo mass function and halo-size distributions derived from kinematic and lensing data with results from different cosmological simulations.

\begin{acknowledgements}
        AA was supported by a grant from VILLUM FONDEN (project number 16599). This project is partially funded by PRIM-MIUR 2017WSCC32 ``Zooming into dark matter and proto-galaxies with massive lensing clusters'' and the Danish council for independent research under the project ``Fundamentals of Dark Matter Structures'', DFF--6108-00470. MM acknowledges support from the Italian Space Agency (ASI) through contract ``Euclid - Phase D" and from the grant MIUR PRIN 2015 ”Cosmology and Fundamental Physics: illuminating the Dark Universe with Euclid”. CG acknowledges support by VILLUM FONDEN Young Investigator Programme through grant no. 10123.
\end{acknowledgements}

\bibliographystyle{aa}
\bibliography{bibliography}

\clearpage

\appendix
\section{pPXF spectral simulations}
\label{sec.:appendix_ppxf}
\subsection{Simulated sample}
We describe here our simulations of galaxy spectra to assess the robustness of the velocity dispersion measurements with the pPXF software \citep{Cappellari_2017} which are critical in our lens model analysis. A sample of 10000 simulated spectra of early-type galaxies were created according to the following steps: \\
1) From the MACS~1206 MUSE datacube we extracted all the spectra of spectroscopic members with $m_{F160W}\!<\!24$ and \sn > 20, using an aperture of $R=0.8$\arcsec. \\
2) Using pPXF best-fit to the spectra, we determined the stellar templates and relative weights, whose linear combination reproduce our galaxy spectra, which include younger stellar populations in some cases with significant Balmer lines. From this set of template, we created rest-frame synthetic model spectra that is not yet convoluted with a LOSVD and without noise, with a dispersion of 0.4\,\AA\ per pixel and a spectral resolution of 1.35\,\AA\ FWHM. \\
3) Model spectra are convoluted with a LOSVD up to the $4^{th}$ moment, with velocity dispersion ranging from 0 to $ 250$\,\vel. \\
4) We assigned to each model spectrum a redshift so to reproduce the observed redshift distribution of cluster galaxies. The spectral resolution is then degraded to the lower MUSE instrumental resolution of 2.6\,\AA\ FWHM approximately constant over the whole considered wavelength range.\\
6) The spectra wavelength-range is matched to the MUSE range and re-sampled to MUSE 1.25\,\AA/pix scale. Gaussian noise, drawn from the MUSE variance spectra, is added to the spectra in such a way to reproduce a \sn from $5$ to $100$. In this way, the high variance due to sky subtraction around sky lines is included in the simulations. \\
The final set of simulations is divided into six subsets that differ for their input $\sigma$ and \sn (see Table\,\ref{table:simulation_parameters}). 
The first ($V_{in}$), third ($h_3^{in}$), and fourth ($h_4^{in}$) moments of the LOSVD are uniformly distributed within the intervals ($-50.0,\,  50.0$)\,\vel, ($-0.1,\, 0.1$), and ($-0.1,\, 0.1$), respectively. 

\begin{table}[b!]  
	\label{table:simulation_parameters}    
	\def\arraystretch{1.5}
	\centering          
	\begin{tabular}{|c | c | c|}
		\hline
		\boldmath{N} & \boldmath{$\sigma_{in}\, \mathrm{[km\ s^{-1}]}$} & \boldmath{\sn}\\ 
		\hline
		\hline
		$1000$ & $0.10$ - $250.0$  & $5.0$ - $100.0$ \\  
		\hline
		$1000$ & $10.0$ - $150.0$ & $5.0$ - $30.0$ \\
		\hline
		$1000$ & $10.0$ - $150.0$ & $5.0$ - $15.0$ \\
		\hline
		$2000$ & $45.0$ - $250.0$ & $5.0$ - $100.0$ \\
		\hline
		$2000$ & $45.0$ - $250.0$ & $5.0$ - $20.0$ \\
		\hline
		$3000$ & $50.0$ - $250.0$ & $5.0$ - $15.0$ \\
		\hline                  
	\end{tabular}
	\smallskip
    \caption{Number of simulated spectra of cluster members in six bins of velocity dispersion, $\sigma_{in}$, and \sn. In each subset, LOSVD moments ($V_{in}$, $h_3^{in}$, $h_4^{in}$) are randomly distributed (see text).}    
\end{table}
\begin{figure}
    \centering
    \includegraphics[width=9cm]{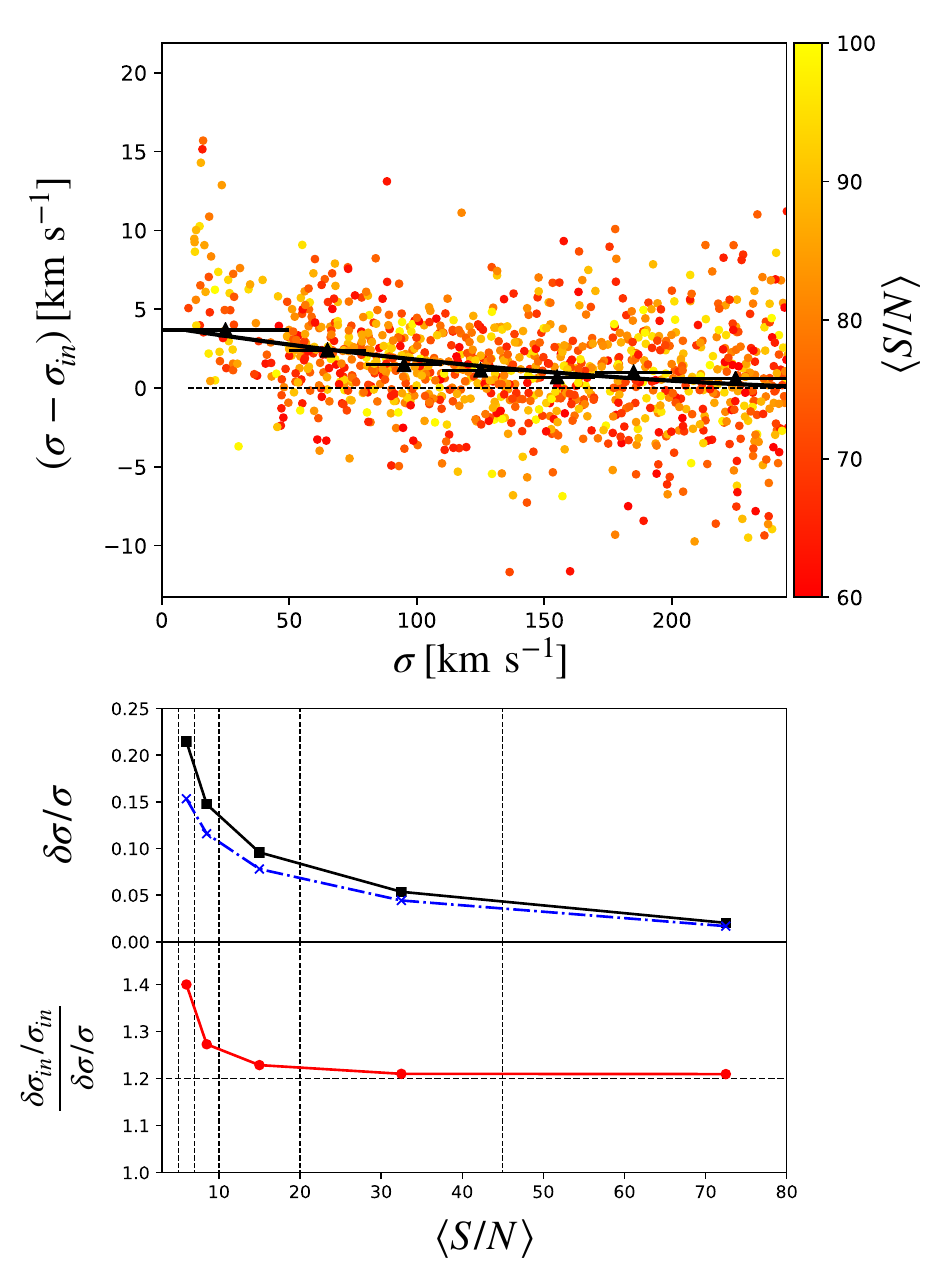}
    \caption{\textit{Top:} Difference between pPXF measured velocity dispersions ($\sigma$) and input velocity dispersions in simulated spectra ($\sigma_{in}$), as a function of $\sigma$. Only the galaxies with $\left<S/N\right>>60$ are shown here. The data points are color shaded according to the spectra \sn. Black triangles are the medians of the distributions in 7 different $\sigma$ bins. The black polynomial line underlines the small increasing bias at low $\sigma$. \newline \textit{Bottom:} The black solid (blue dashed) line shows the relative statistical error estimated from simulations (provided by pPXF), as a function of \sn. The red curve corresponds to the ratio between the two; the statistical relative errors appear to be underestimated by $\sim\! 20\%$ in our spectra.}
    \label{fig:ppxf}
\end{figure}

First, we investigated the possible presence of systematics, especially at low \sn and $\sigma\lesssim 100$\,\vel, where the MUSE instrumental resolution may affect our measurements. Fig.\,\ref{fig:ppxf} shows that an increasing bias in velocity dispersion measurements at low velocities is detectable in the high \sn regime ($\gg 60$). Using  a polynomial function to fit the median values of the $\sigma-\sigma_{in}$ distributions in seven $\sigma$ bins, we found a correction for measured velocity dispersions, $\sigma$, given by: 

\begin{equation}
    \sigma_{true} = \sigma-(4.00 \cdot 10^{-5}\sigma^2 -2.59 \cdot 10^{-2}\sigma + 4.06)\, \mathrm{ km\, s^{-1}}.
\label{eq: sigma_cor}
\end{equation}

The increase in the $\sigma-\sigma_{in}$ scatter at low $\left<S/N\right>$ makes it difficult to detect such a small bias at lower \sn.
The scatter in $\sigma-\sigma_{in}$ increases significantly below \sn$=10$, with relative errors exceeding 10\% and measurements become increasingly sensitive to the presence of sky lines. For this reason, we chose a limiting \sn of 10 in our analysis. \\
Another goal of the simulations was to quantify realistic errors on velocity dispersion measured by pPXF. In our case, we found that relative errors, $\delta\sigma/\sigma$, are systematically underestimated by $\sim\!20\%$ for \sn$\gg\!15$ and of $\sim\!25\%$ for \sn$\sim\!10$. 
In all our measurements, velocity dispersions are corrected for the small bias in the high \sn regime (Eq.\,\ref{eq: sigma_cor}) and statistical errors are corrected using a polynomial fit, as shown in Fig.\,\ref{fig:ppxf}/bottom.

\clearpage

\section{$\sigma\,\mbox{-}\,L$ scaling relation from kinematic measurements}
\label{sec.:appendix_emcee}
Given a set of N cluster members of magnitude $m_i^{gal}$ with measured velocity dispersions, $\sigma^{gal}_{ap,i} \pm \delta\sigma^{gal}_{ap,i}$, the $\sigma\,\mbox{-}\,mag$ scaling relation, corresponding to the $\sigma\,\mbox{-}\,L^\alpha$ relation in Eq.\,\ref{eq.: Scaling_relation_sigma},  can be written as:

\begin{equation}
    \label{eq.: sigma_mag}
    \hat{\sigma}_{ap,i}^{gal}=\sigma_{ap}^{ref} 10^{\,0.4\, (m^{ref}_{F160W}-m_i^{gal})\,\alpha},
\end{equation}

\null\noindent where $\hat{\sigma}_{ap,i}^{gal}$ are the model predicted velocity dispersions for a cluster member with F160W magnitude $m_i^{gal}$.\\
To estimate the  model parameters of the scaling relation and their uncertainties, we sample the posterior distribution of $\sigma_{ap}^{ref}$ and $\alpha$, including the intrinsic scatter  $\Delta \sigma_{ap}$ of the measured velocities around the backbone of the scaling relation. Using Bayes' theorem, the posterior probability function can be written as: 

\begin{multline}
    \label{eq.: posterior}
    p\left(\sigma_{ap}^{ref}, \alpha, \Delta \sigma_{ap} \mid m^{gal}, \sigma^{gal}_{ap}, \delta\sigma^{gal}_{ap}\right) \propto \\ p\left(\sigma^{gal}_{ap} \mid m^{gal}, \delta\sigma^{gal}_{ap}, \sigma^{ref}_{ap}, \alpha, \Delta \sigma_{ap}\right) p\left(\sigma_{ap}^{ref}, \alpha, \Delta \sigma_{ap}\right).
\end{multline}

In particular, the posterior is the product of a likelihood function (\ref{eq.: likelihood}) and a prior (\ref{eq.: prior}):

\begin{multline}
    \label{eq.: likelihood}
    \ln \left[p\left(\sigma_{ap}^{gal} \mid m^{gal}, \delta\sigma_{ap}^{gal}, \sigma_{ap}^{ref}, \alpha, \Delta \sigma_{ap}\right)\right] =\\= - \frac{1}{2} \sum_{i=1}^N \left\{ \frac{\left(\sigma_{ap,i}^{gal}-\hat{\sigma}_{ap,i}^{gal}\right)^2}{\left(\delta\sigma^{gal}_{ap,i}\right)^2 + \Delta \sigma_{ap}^2} + \ln \left[2 \pi \left(\left(\delta\sigma^{gal}_{ap,i}\right)^2 + \Delta \sigma_{ap}^2 \right) \right] \right\},
\end{multline}

\begin{equation}
    \label{eq.: prior}
    \ln\left[ p\left(\sigma_{ap}^{ref}, \alpha, \Delta \sigma_{ap}\right)\right]=
    \begin{cases}
        - \ln(\Delta \sigma_{ap}),& \scriptstyle \text{if } \sigma_{min}^{ref}<\sigma_{ap}^{ref}<\sigma_{max}^{ref}\\ & \scriptstyle \text{and } \alpha_{min}<\alpha<\alpha_{max} \\ & \scriptstyle \text{and } \left(\Delta \sigma_{ap}\right)_{min}<\Delta \sigma_{ap}<\left(\Delta \sigma_{ap}\right)_{max}\\
        -\infty,              & \scriptstyle \text{otherwise}
    \end{cases}.
\end{equation}
The boundaries $\sigma_{min}^{ref}$, $\sigma_{max}^{ref}$, $\alpha_{min}$, $\alpha_{max}$, $\left(\Delta \sigma_{ap}\right)_{min}$ and $\left(\Delta \sigma_{ap}\right)_{max}$ were chosen to limit the parameter space around the measured velocity dispersions.

To sample the log-posterior in the 3D parameters space ($\sigma_{ap}^{ref}$, $\alpha$, $\Delta \sigma_{ap}$), we use the Affine-Invariant Markov Chain Monte Carlo (MCMC) Ensemble sampler developed by Goodman and Weare \citep{gw2010}, and in particular its \texttt{python} implementation\footnote{https://emcee.readthedocs.io/en/latest/} \citep{emcee_2013}. The parameter space is explored with $100$ walkers, with $5000$ steps each, which are initialized in a narrow Gaussian sphere around the maximum-likelihood point.
To ensure that the final distributions are independent from the initial walker positions, we remove 80 steps in the burn-in phase based on the auto-correlation time computed for each parameter.

\begin{figure}
	\centering
	\includegraphics[width=9cm]{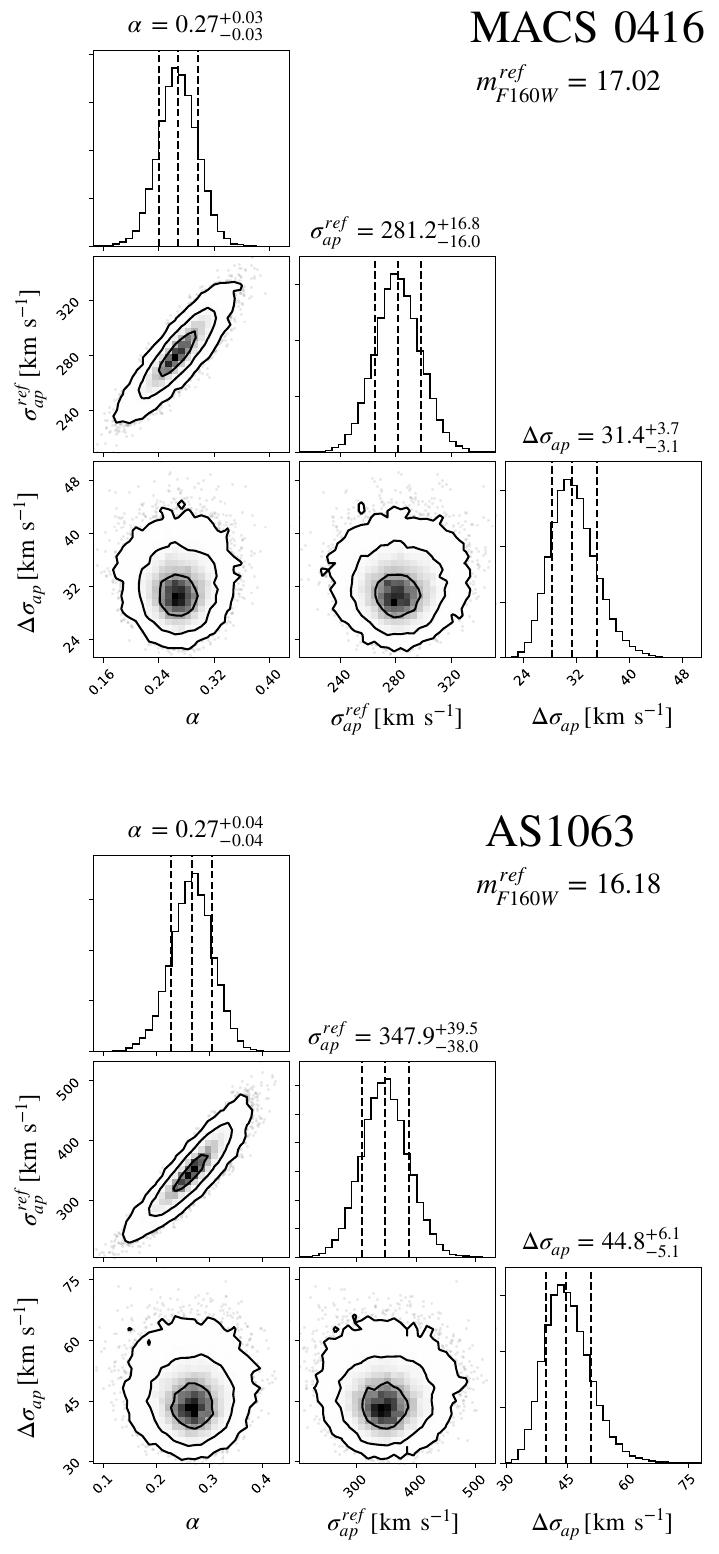}
	\caption{Posterior probability distributions for scaling relation parameters in Eq.\,\ref{eq.: sigma_mag} obtained from measured velocity dispersions of cluster members in \mb\ and \ab. The $16^{th}$, $50^{th}$ and $84^{th}$ percentiles of the marginalized distributions for the slope ($\alpha$), normalization ($\sigma_{ap}^{ref}$) and scatter around the scaling relation ($\Delta \sigma_{ap}$) are quoted and shown as dashed lines in each histogram.}
	\label{fig:emcee_two}
\end{figure}

The posterior probability distributions of model parameters (Eq.\,\ref{eq.: posterior}) obtained with this procedure are shown in Fig.\,\ref{fig:emcee_macs1206} for \ma\ and Fig.\,\ref{fig:emcee_two} for \mb\ and \ab.

\clearpage

\section{dPIE mass distribution}
\label{sec.:appendix_dPIE_mass}

The dual pseudo-isothermal elliptical mass distribution (dPIE), described in \cite{Eliasdottir_lenstool} and \cite{Limousin_lenstool}, is defined by eight parameters; three for the sky position and redshift ($z$), two for ellipticity ($e$) and position angle, and three for core radius ($r_{core}$), truncation radius ($r_{cut}$) and central velocity dispersion ($\sigma_{0}$).
All the halos in our lens models are described as dPIEs profiles, cluster-scale halos are generally included in the limit $r_{cut}\rightarrow\infty$; in this case, the dPIE reduces to a pseudo isothermal elliptical mass distribution (PIEMD), defined by \cite{Kassiola_1993}. Sub-halos associated to cluster members are instead parametrized as {\sl circular}  dPIEs of vanishing $r_{core}$, with $r_{cut}$ and $\sigma_0$ following the scaling relations in Eq.\,\ref{eq.: Scaling_relation_sigma} and \ref{eq.: Scaling_relation_rcut} (see Sec.~\ref{sec.:lensing_models}).\\
The equations below, mostly derived in \cite{Eliasdottir_lenstool} and \cite{Limousin_lenstool}, refer to circular ($e=0$) dPIEs. For completeness, we give the expressions for projected masses and surface-mass densities and show how the dPIE $\sigma_0$ parameter can be derived from the aperture average line-of-sight (projected) velocity dispersion $\sigma_{ap}$, which is associated to our kinematic measurements.

\subsection{dPIE density and mass}
The mass density profile for the spherical dPIE, as implemented in {\texttt{LensTool}}, is defined through the 3D-density \citep{Limousin_lenstool}:

\begin{gather}
    \label{eq.: Density_3d_dPIE}
    \rho(r) =  \frac{\rho_0}{(1+r^2/r^2_{core})(1+r^2/r^2_{cut})},
\end{gather}

\noindent where $r$ is the distance from the center of the mass distribution, while $r_{core}$ and $r_{cut}$ are the core and truncation radii respectively (with $r_{cut}>r_{core}$). Eq.\,\ref{eq.: Density_3d_dPIE} shows a smooth separation between an isothermal behaviour, $\rho \propto r^{-2}$, where $r_{core}<r<r_{cut}$ and a steeper decrease, $\rho \propto r^{-4}$, for $r>r_{cut}$.\\
The relation between the central density $\rho_0$ and the 1D-central velocity dispersion, $\sigma_{0}$, is \citep{Limousin_lenstool}:

\begin{gather}
    \label{eq.: Central_density_dPIE}
    \rho_0 =  \frac{\sigma^2_0}{2 \pi G} \frac{r_{cut}+r_{core}}{r_{core}^2 r_{cut}}.
\end{gather}

\noindent In the Singular Isothermal Sphere (SIS) limit, $r_{core}\rightarrow 0$ and $r_{cut}\rightarrow \infty$, Eq.\,\ref{eq.: Density_3d_dPIE} reduces to the familiar expression $\rho_{SIS}=\sigma_0^2/(2\pi G\, r^2)$. 
In {\texttt{LensTool}}, dPIEs are implemented trough a fiducial velocity dispersion $\sigma_{LT}$ related to the central velocity dispersion $\sigma_0$ by: $\sigma_0=\sqrt{3/2}\, \sigma_{LT}$. \\
Projecting the 3D-density profile on a plane perpendicular to the line-of-sight, we obtain the projected dPIE surface-mass-density as a function of the projected distance from the center $R$ \citep{Eliasdottir_lenstool}:

\begin{eqnarray}
    \nonumber \Sigma(R) & = & 2 \int_{R}^{\infty} \frac{\rho(r)r}{\sqrt{r^2-R^2}}dr =\\
    & = & \frac{\sigma_0^2}{2G}\frac{r_{cut}}{r_{cut}-r_{core}} \left(\frac{1}{\sqrt{r_{core}^2+R^2}}-\frac{1}{\sqrt{r_{cut}^2+R^2}}\right).
    \label{eq.: Density_2d_dPIE}
\end{eqnarray}

\begin{figure}[h]
	\centering
	\includegraphics[width=9cm]{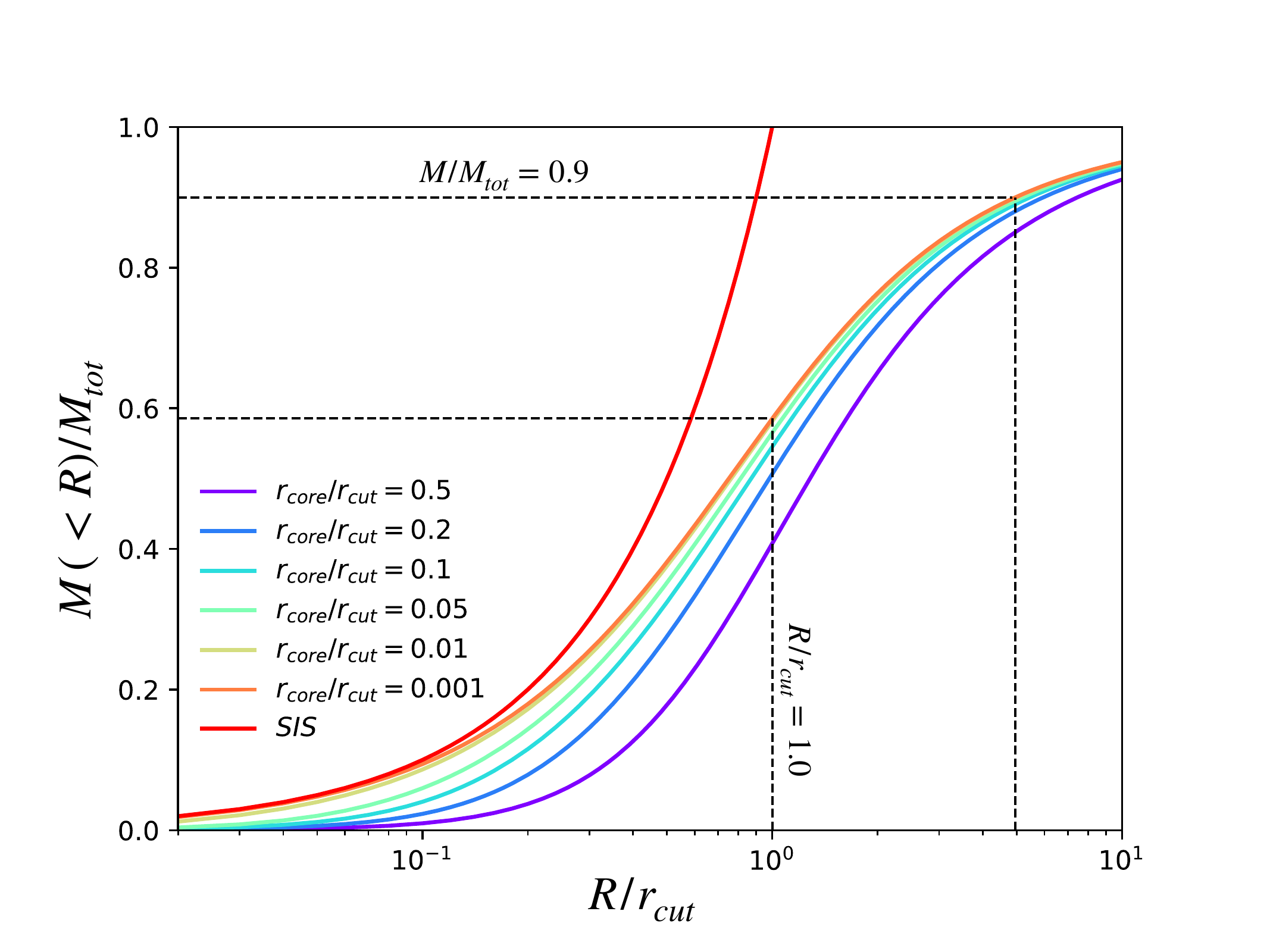}
	\caption{Fraction of projected mass over total mass as a function of aperture radii (in units of $r_{cut}$) for dPIE profiles with different $r_{core}/r_{cut}$ values. For small $r_{core}/r_{cut}$,    $\sim\! 60\%$ of the total mass is contained within $r_{cut}$, while the $90\%$ is contained within $5r_{cut}$.}
	\label{fig:dPIE_profiles}
\end{figure}

Integrating Eq.\,\ref{eq.: Density_3d_dPIE}, we can also write the total 3D mass $m(r)$ enclosed by a sphere of radius $r$:

\begin{multline}
    \label{eq.: Mass_3d_dPIE}
    m(r) = 4 \pi \int_{0}^{r} \rho(r')r'^2 dr' =\\ =\frac{2\sigma_0^2}{G}\frac{r_{cut}}{r_{cut}-r_{core}} \left[r_{cut} \arctan \left(\frac{r}{r_{cut}}\right)-r_{core} \arctan \left(\frac{r}{r_{core}}\right)\right].
\end{multline}

Similarly, the total projected mass within an aperture of projected radius $R$, $M(R)$, is (from Eq.\,\ref{eq.: Density_2d_dPIE}):

\begin{multline}
    \label{eq.: Mass_2d_dPIE}
    M(R) = 2 \pi \int_{0}^{R} \Sigma(R')R' dR' =\\ =\frac{\pi \sigma_0^2}{G}\frac{r_{cut}}{r_{cut}-r_{core}} \left(\sqrt{r_{core}^2+R^2}-r_{core}-\sqrt{r_{cut}^2+R^2}+r_{cut}\right).
\end{multline}

The dPIE profile, contrary to the SIS, has a finite total mass $M_{tot}$ whose value can be obtained from Eq.\,\ref{eq.: Mass_3d_dPIE} (or Eq.\,\ref{eq.: Mass_2d_dPIE}) in the limit of r$\rightarrow\infty$ (R$\rightarrow\infty$):

\begin{gather}
    \label{eq.: Total_mass_dPIE}
    M_{tot} = \frac{\pi \sigma_0^2 r_{cut}}{G}.
\end{gather}

In the limit $r_{core}/r_{cut}\!\rightarrow\!0$, a sphere of radius $r=r_{cut}$ encloses  half of the total 3D dPIE mass. While 60 (90)\% of the total projected mass is included within a projected radius of $R=r_{cut}$ ($R=5r_{cut})$ (see Fig.\,\ref{fig:dPIE_profiles}). \\
Similarly, \cite{Eliasdottir_lenstool} give an expression for the half-mass radius, $R_{M_{tot}/2}$, that is enclosing half of the total projected mass, from Eq.\,\ref{eq.: Mass_2d_dPIE}:

\begin{gather}
    \label{eq.: Half_mass_r}
    R_{M_{tot}/2} = \frac{3}{2}\sqrt{r_{core}^2+\frac{10}{3}r_{core}r_{cut}+r_{cut}^2}\ .
\end{gather}
Finally, from Eq.\,\ref{eq.: Mass_3d_dPIE} we derive an expression for the circular velocity $v_c(r)=\sqrt{G\ m(r)/r}$:

\begin{gather}
    \label{eq.: circular_velocity}
    v_c^2(r) = 2\sigma_0^2\frac{r^2_{cut}}{r_{cut}-r_{core}} \frac{1}{r} \left[ \arctan \left(\frac{r}{r_{cut}}\right)-\frac{r_{core}}{r_{cut}} \arctan \left(\frac{r}{r_{core}}\right)\right],
\end{gather}

\noindent which has a maximum at $v_c^{max}$. In the SIS limit, Eq.\,\ref{eq.: circular_velocity} reduces to $v_c^{SIS}=\sqrt{2}\,\sigma_0$. To compare the circular velocity $v_c^{max}$ in \cite{Grillo_2015} and \bo\ with our sigma functions in Fig.\,\ref{fig:velocity_distribution}, we adopt $\sigma_0=v_c^{max}/\!\sqrt{2}$, as one  can easily verify that for the full range of $r_{core}$ and $r_{cut}$ values of our member galaxies, $v_c^{max}$ is smaller than $v_c^{SIS}$ by less than the $3\%$.

\subsection{dPIE velocity dispersion}
\label{sec.:appendix_velocity_dispersion}
Assuming an isotropic mass distribution, the observed line-of-sight velocity dispersion at a projected distance $R$  from the center is given by (\citealt{Eliasdottir_lenstool}, \citealt{Agnello_2014}, \citealt{Binney_Mamon_1982}):

\begin{gather}
    \label{eq.: sigma_los}
    \sigma_{los}^2(R) = \frac{2 G}{I(R)} \int_{R}^{\infty} \frac{m(r) \nu(r)}{r^2}\sqrt{r^2-R^2}dr ,
\end{gather}

\noindent where $\nu(r)$ is the 3D-luminosity density of the galaxy which is related to the surface brightness profile, $I(R)$, by:
\begin{gather}
    \label{eq.: luminosity_density}
    \nu(r) = -\frac{1}{\pi}\int_{r}^{\infty} \frac{\partial_R(I(R))}{\sqrt{R^2 -r^2}}dR .
\end{gather}

Similarly, the average line-of-sight velocity dispersion inside an aperture radius R, $\sigma_{ap}$, for an isotropic system, is given by:

\begin{gather}
    \label{eq.: Sigma_aperture_dPIE}
    \sigma_{ap}^2(R) = \frac{2 \pi}{L(R)} \int_{0}^{R} \sigma_{los}^2(R') I(R')R' dR',
\end{gather}

\noindent where $L(R)$ is the total projected luminosity within $R$:

\begin{gather}
    \label{eq.: total_projected_luminosity}
    L(R)=2 \pi \int_{0}^{R} R' I(R') dR'.
\end{gather}

$\sigma_{ap}$ is the quantity that we can directly compare to our kinematic measurements (see Sec.~\ref{sec.:lensing_models} and \ref{sec.:Lensing_kinematics}) and it can also be rewritten in terms of the measurable surface brightness of the elliptical galaxy as \citep{Agnello_2014}:

\begin{eqnarray}
    \nonumber
    \sigma_{ap}^2(R) & = &\frac{4G}{3L(R)} \left[\int_{0}^{\infty}R'I(R')\int_{0}^{R'}\frac{4 \pi \rho(r) r^2}{\sqrt{R'^2-r^2}}drdR'\right. \\
    \nonumber & - &\left. \int_{R}^{\infty}R'I(R')\int_{R}^{R'}\frac{\partial_r \left(m(r)(r^2-R^2)^{3/2}/r^3\right)}{\sqrt{R'^2-r^2}}dr dR'\right]\\
    & \equiv & \frac{2}{3}\sigma_0^2 c_p^2(R).
    \label{eq.: sigma_brighteness_dPIE}
\end{eqnarray}

The last line recasts the aperture-averaged $\sigma_{ap}$ in terms of the central velocity dispersion $\sigma_0$ explicitly, with the projection coefficient $c_p(R)$ encompassing all other terms except the prefactor $\sqrt{3/2}.$ The relation between $\sigma_{ap}$ and the {\texttt{LensTool}} fiducial velocity dispersion, $\sigma_{LT}$, can be written as:
\begin{gather}
    \label{eq.: average_aperture_sigma}
    \sigma_{ap}(R)=\sigma_{LT}\, c_p(R).
\end{gather}

\begin{figure}
	\centering
	\includegraphics[width=8.5cm]{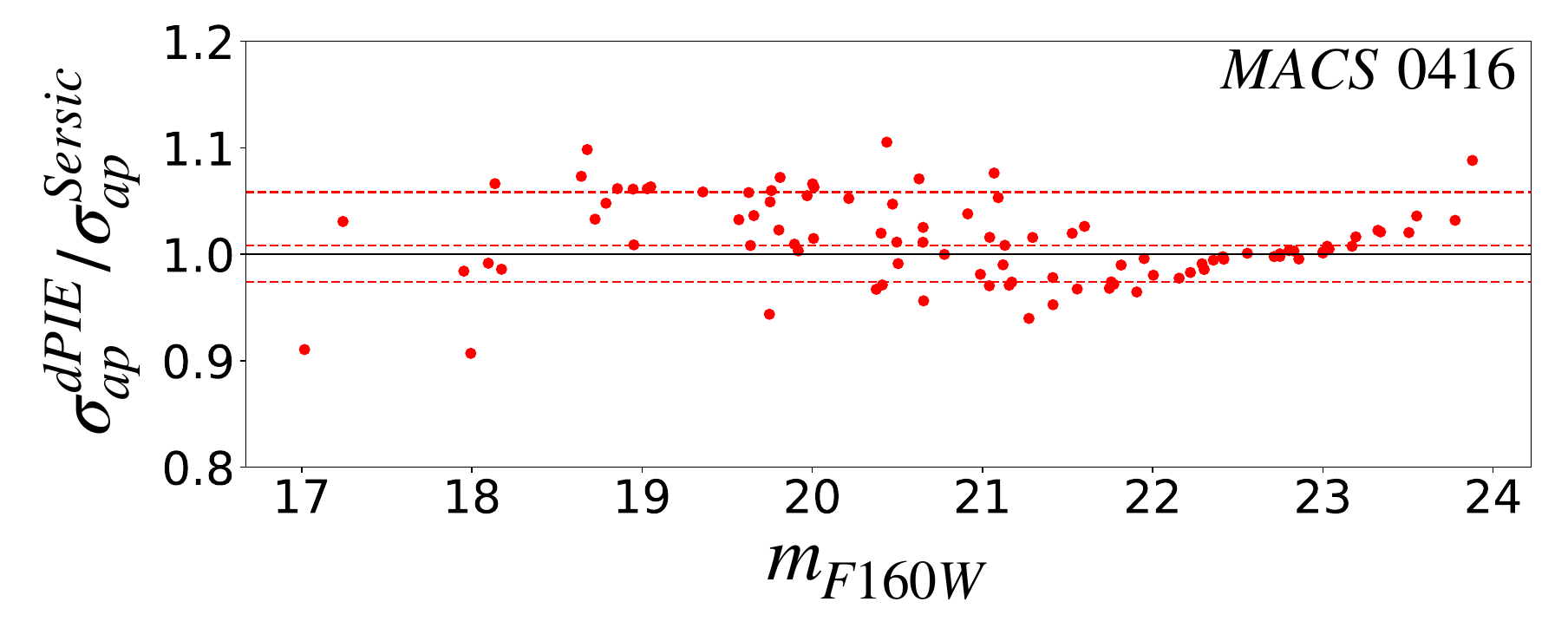}
	\caption{Ratio between aperture average line-of-sight velocity dispersions, computed assuming a dPIE or a Sersic surface brightness profile of 95 cluster members in \mb, as a function of F160W magnitudes. The aperture radius is $R=0.8\arcsec$.  Red dashed lines correspond to the $16^{th}$, $50^{th}$ and $84^{th}$ percentiles of the entire distribution.}
	\label{fig:sersic_dpie}
\end{figure}
\begin{figure}[h!]
	\centering
	\includegraphics[width=9cm]{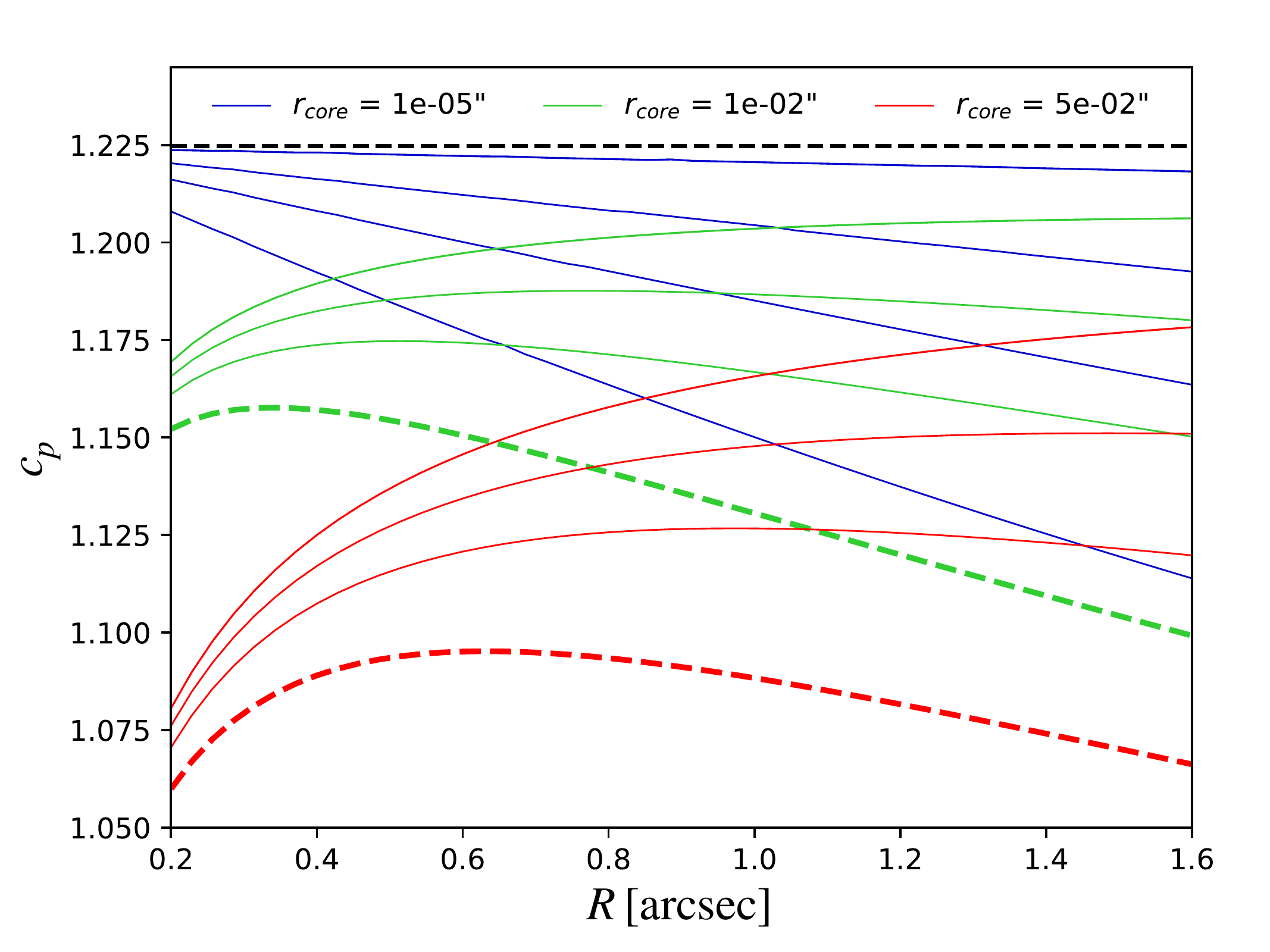}
	\caption{Projection coefficient $c_p$ as a function of aperture radius $R$ in arcseconds. Curves associated to the same $r_{core}$ are grouped with the same color, with $r_{cut}=5,\,10,\,20,\,100$ arcsec, from bottom to top. The thick dashed red and green lines correspond to typical values of $r_{core}$ and $r_{cut}$ derived from our lensing$+$dynamics modeling of clusters. The dashed black horizontal line corresponds to $\sqrt{3/2}$.}
	\label{fig:projection_coefficient}
\end{figure}

To compute $c_p(R)$, a surface brightness profile $I(R)$ for cluster members is needed. For our purpose, we tested two choices for $I(R)$ using: 1) a surface brightness profile scaling with the projected mass density $\Sigma(R)$ (Eq.\,\ref{eq.: Density_2d_dPIE}) derived from the cluster lens model by \bo, and 2) a Sérsic profile extracted from the HST data of \mb\ in the F814 band:

\begin{gather}
    \label{eq.: Sersic}
    I(R)=I_e \exp{\left\{-b_n\left[ \left( \frac{R}{R_e} \right)^{1/n} -1\right] \right\}}.
\end{gather}

The Sérsic index $n$ and the effective radius $R_e$ are determined by fitting the light profile of cluster members using the public software G\tiny ALFIT \normalsize (\citealt{galfit_1} and \citealt{galfit_2}), adopting $b_n= 2n-1/3$ \citep{sersic}.\\
Fig.\,\ref{fig:sersic_dpie} shows that for our spectral aperture radius, $R=0.8\arcsec$, the differences between the $\sigma_{ap}$ obtained assuming a dPIE or a Sérsic surface brightness profile are within $\sim\! 5\%$. For this reason, in our work we always assume a dPIE surface brightness profile which makes the analysis convenient and independent from Sérsic light profile fitting.\\
Assuming $I(R)=\Sigma(R)$, the projection coefficient $c_p(R)$ can be numerically computed as

\begin{multline}
    \label{eq.: projection_coefficient}
    c_p^2(R) = \frac{6}{\pi}\frac{r_{core}+r_{cut}}{r_{core}^2r_{cut}}\left( \sqrt{r_{core}^2+R^2}-r_{core}-\sqrt{r_{cut}^2+R^2}+r_{cut}\right)^{-1} \\
    \cdot \int_{0}^{R}R'\int_{R'}^{\infty} \frac{r_{cut} \arctan \left(\frac{r}{r_{cut}}\right)-r_{core} \arctan \left(\frac{r}{r_{core}}\right)}{(1+r^2/r^2_{core})(1+r^2/r^2_{cut})}\frac{\sqrt{r^2-R'^2}}{r^2}drdR'.
\end{multline}

In Fig.\,\ref{fig:projection_coefficient}, we show the projection coefficients as a function of aperture radius $R$, for different values of  $r_{core}$ and $r_{cut}$. The dashed black line indicates the asynthotic value of $c_p=\sqrt{3/2}$ , which corresponds to the limit $r_{core}\rightarrow 0$ and $r_{cut}\rightarrow\infty$, where the dPIE reduces to a SIS for which $\sigma_0=\sqrt{3/2}\,\sigma_{LT}$.

\end{document}